\documentclass{aastex63}
\usepackage{verbatim}
\usepackage{float}
\usepackage{graphicx}
\usepackage{epstopdf}
\usepackage{subfigure}
\usepackage{CJK}

\usepackage{booktabs}
\usepackage{url}
\usepackage{amssymb}
\makeatletter

\newcommand{\Rmnum}[1]{\expandafter\@slowromancap\romannumeral #1@}
\makeatother

\shorttitle{}
\shortauthors{}

\begin{document}
\begin{CJK*}{UTF8}{gbsn}

\title{Characterizing the $\gamma$-Ray Variability of Active Galactic Nuclei with Stochastic Process Method}

\correspondingauthor{Dahai Yan}
\email{yandahai@ynao.ac.cn}
\correspondingauthor{Li Zhang}
\email{lizhang@ynu.edu.cn}

\author{Haiyun Zhang (张海云)}
\affiliation{Department of Astronomy, Key Laboratory of Astroparticle Physics of Yunnan Province, Yunnan University, \\Kunming 650091, China}

\author{Dahai Yan (闫大海)}
\affiliation{Key Laboratory for the Structure and Evolution of Celestial Objects, Yunnan Observatories, Chinese Academy of Sciences,\\
	Kunming 650011, China}

\author{Li Zhang (张力)}
\affiliation{Department of Astronomy, Key Laboratory of Astroparticle Physics of Yunnan Province, Yunnan University, \\Kunming 650091, China}

\begin{abstract}
The $\gamma$-ray astronomy in time domain has been by now progressed further as the variabilities of Active Galactic Nuclei (AGNs) on different timescales
have been reported a lot. We study the $\gamma$-ray variabilities of 23 jetted AGNs through applying a stochastic process method to the $\sim$12.7 yr long-term light curve (LC) obtained by Fermi-Large Area Telescope (Fermi-LAT). 
In this method,
the stochastically driven damped simple harmonic oscillator (SHO) and the damped random walk (DRW) models are used to model the long-term LCs. 
Our results show that the long-term variabilities of 23 AGNs can be characterized well by both SHO and DRW models.
However, the SHO model is restricted in the over-damped mode and the parameters are poorly constrained.
The SHO power spectral densities (PSDs) are same as the typical DRW PSD. 
In the plot of the rest-frame timescale that corresponds to the broken frequency in the PSD versus black hole mass, the intrinsic $\gamma$-ray characteristic timescales of 23 AGNs occupy almost the same space with the optical variability timescales obtained from the accretion disk emission.
This suggests a connection between the jet and the accretion disk. Same as the optical variability of AGN accretion disk, the $\gamma$-ray timescale is also consistent with the thermal timescale caused by the thermal instability in the standard accretion disk of AGN. 
\end{abstract}


\section{Introduction} \label{sec:intro}

High energy (HE; $\geqslant$ 100 MeV) $\gamma$-ray observations suggest that the emissions from Active Galactic Nuclei (AGNs) 
dominate the extragalactic $\gamma$-ray sky \citep{2020ApJS..247...33A}. The strongly Doppler-boosted blazars, an extreme class of AGNs of which emissions are mainly from the nonthermal relativistic jets, are dominant in these powerful emitters. Blazars are classified into BL Lac objects (BL Lacs) and flat spectrum radio quasars (FSRQs), according to the strength of their optical emission lines \citep{2020ApJ...892..105A}. FSRQs have strong, broad emission lines, while BL Lacs have weak, narrow, or no such lines.

AGN variability has already been detected at entire electromagnetic wavelengths with timescales covering from decades down to minutes. 
Radio-loud AGNs are highly variable $\gamma$-ray emitters. This not only applies to the blazars which have strong and incessant flux variability, 
but also to misaligned jet sources such as radio galaxies \citep{2018A&A...617A..91M,2019A&A...623A...2A}. 
The underlying physical processes can be investigated by characterizing the variabilities \citep[e.g.,][]{2017A&A...598A..39H,2018ApJ...864..164Y,2019Galax...7...28R,2020ApJ...891..120B}.
 
The observations of Fermi-Large Area Telescope (Fermi-LAT) have advanced the studies in HE time-domain.
An attractive phenomenon is the $\gamma$-ray quasi-periodic oscillations (QPOs)
detected in LAT data of blazars \cite[e.g.,][]{2015ApJ...813L..41A,2016AJ....151...54S,2018NatCo...9.4599Z,2020ApJ...896..134P,2020ApJ...891..163Z}. 
However, the reliability of these QPOs is always questionable \citep[e.g.,][]{2019MNRAS.482.1270C,2020A&A...634A.120A,2021ApJ...907..105Y}.
Another interesting phenomena is the fast $\gamma$-ray flares on timescale of a few minutes detected in the LAT data of FSRQs \citep{2016ApJ...824L..20A,2019ApJ...877...39M,2020NatCo..11.4176S}.
Besides, statistical characteristics of the $\gamma$-ray variability in AGNs have been extensively investigated.
The commonly used methods are the analyses of power spectral density (PSD) and flux distribution \citep[e.g.,][]{2010ApJ...722..520A,2018RAA....18..141S,2019ApJ...877...39M}.
\cite{2010ApJ...722..520A} presented $\gamma$-ray variabilities of the 106 AGNs systematically, 
using the first 11 months of the Fermi survey. 
They reported that more than 50 $\%$ of the sources are found to be variable with a power-law (PL) PSD, 
and a random walk underlying mechanism was reflected in some blazars.
\cite{2013ApJ...773..177N} reported a characteristic timescale of $\sim$7.9 days in the PSD of 3C 454.3 by analyzing the first four years Fermi-LAT data, and they used an internal shock model to interpret this timescale. 
\citet{2019ApJ...877...39M} presented a detailed analysis of  LAT LCs of bright $\gamma$-ray FSRQs, and put strong constraints on blazar jet physics accordingly.

A stochastic process model has been wildly used to describe optical variability of AGN accretion disk \citep[e.g.,][]{2009ApJ...698..895K,2010ApJ...721.1014M,2013ApJ...765..106Z,2017ApJ...842...96R,2018MNRAS.476L..55L,2018ApJ...853..116Z,2021Sci...373..789B}. Generally, the damped random walk (DRW) model can provide successful fit to the long-term variability of AGN accretion disk.
It is proved that such a stochastic process model provides a powerful tool to extract information from AGN variability \citep[e.g.,][]{2017MNRAS.470.3027K,2021Sci...373..789B}.
In the past few years, the stochastic process model has been applied to $\gamma$-ray variabilities of AGNs in several papers 
\citep{2014ApJ...786..143S,2018ApJ...863..175G,2019ApJ...885...12R,2020ApJS..250....1T,2020ApJ...895..122C,2021ApJ...907..105Y,2021ApJ...919...58Z}.
Based on the stochastic model developed by \cite{2009ApJ...698..895K}, \cite{2014ApJ...786..143S} modeled the $\gamma$-ray LCs of 13 blazars observed 
during the first four years of the Fermi sky survey with the Ornstein–Uhlenbeck (OU) process (also called DRW) and a linear superposition of the OU processes (sup-OU). They showed that 10 of 13 blazars prefer the sup-OU process over the OU process.
The continuous-time autoregressive moving average (CARMA) method \citep{2014ApJ...788...33K}, a generalized stochastic model which can be applied to astronomical time series, is flexible to capture the features of flux variability and to produce more accurate PSD. 
Applying this method to the 9.5 yr LAT data of the same 13 blazars in \cite{2014ApJ...786..143S},  \cite{2019ApJ...885...12R} reported that the DRW model is good enough to describe the $\gamma$-ray variability of the 13 blazars.
In addition to CARMA, {\it celerite} is a newly developed method for modeling LC with the stochastic process model  \citep{2017AJ....154..220F}.
It was applied to the LAT data of AGNs to examine the significance of the $\gamma$-ray QPOs \citep{2021ApJ...907..105Y,2021ApJ...919...58Z}.

So far, the Fermi-LAT with collecting data for more than 12 yr 
has provided an excellent opportunity to study the long-term $\gamma$-ray variability in AGNs. 
In this paper, we apply the {\it celerite} model to 12.7 yr Fermi-LAT LCs of 23 bright LAT AGNs including 10 BL Lac objects, 12 FSRQ objects and one radio galaxy. We aim to investigate the $\gamma$-ray variabilities of AGNs on long-term timescales.
The format of this paper is as follows.
In Section \ref{sec:style}, we briefly introduce the Fermi-LAT data processing method. 
In Section \ref{sec:model}, the stochastic process models are briefly described. 
In Section \ref{sec:results}, we show the modeling results for the LCs of 23 AGNs with {\it celerite} method. 
In Section \ref{sec:tau-mass fit}, we focus on the variability characteristic timescales in the jets, and compare them with optical results obtained from AGN accretion disk emissions. 
In Section \ref{sec:Discussion}, we discuss our results. Finally, a summary is presented in the Section \ref{sec:Summary}.  

\begin{deluxetable*}{ccccccccc}
	\tablecaption{Information of 23 AGNs.\label{tab:information}}
	\tablewidth{0pt}
	\setlength{\tabcolsep}{2mm}{
	\tablehead{
		\colhead{4FGL Name} & \colhead{R.A.} & \colhead{Decl.} & \colhead{Identification} & \colhead{$z$} & \colhead{Type} & \colhead{log (Eddington ratio)} & \colhead{$\rm {log}\ (M_{\rm BH}/M_{\rm \odot}$}) & \colhead{Ref.} 
	}
	\decimalcolnumbers
	\startdata
	4FGL J2253.9+1609 & 343.4963 & 16.1506 & 3C 454.3 & 0.859 & FSRQ & -0.354 & $9.1\pm0.5$ & 1 \\
	4FGL J1256.1-0547 & 194.0415 & -5.7887 & 3C 279 & 0.5362 & FSRQ & -1.121 & $8.5\pm0.5$ & 1  \\
	4FGL J1427.9-4206 & 216.9866 & -42.106 & PKS 1424-41 & 1.522 & FSRQ & $\cdots$ & $9.2\pm0.5$ & 2 \\
	4FGL J1512.8-0906 & 228.2147 & -9.1064 & PKS 1510-089 & 0.36 & FSRQ & -0.726 & $8.6\pm0.5$ & 1  \\
	4FGL J1504.4+1029 & 226.1033 & 10.4978 & PKS 1502+106 & 1.8378 & FSRQ & -0.986 & $8.7\pm0.5$ & 2 \\
	4FGL J2202.7+4216 & 330.6946 & 42.2821 & BL Lac & 0.0686 & BLL & -2.805 & $8.5\pm0.2$ & 1 \\
	4FGL J2158.8-3013 & 329.7141 & -30.2251 & PKS 2155-304 & 0.1167 & BLL & $\cdots$ & 8.9 & 3 \\
	4FGL J0538.8-4405 & 84.7089 & -44.0862 & PKS 0537-441 & 0.894 & BLL & -0.749 & $8.7\pm0.5$ & 1 \\
	4FGL J0222.6+4302 & 35.6696 & 43.0357 & 3C 66A & 0.37 & BLL & $\cdots$ & $8.57^{+0.03}_{-0.6}$ & 4,12 \\
	4FGL J0319.8+4130 & 49.9579 & 41.5121 & NGC 1275 & 0.01756 & RDG & $\cdots$ & $7.2\pm0.5$ & 5 \\
	4FGL J2232.6+1143 & 338.1525 & 11.7306 & CTA 102 & 1.032 & FSRQ & -0.149 & 8.7 & 6 \\
	4FGL J0457.0-2324 & 74.2608 & -23.4149 & PKS 0454-234 & 1.003 & FSRQ & $\cdots$ & $9.2\pm0.5$ & 7 \\
	4FGL J1224.9+2122 & 186.2277 & 21.3814 & 4C +21.35 & 0.43383 & FSRQ & -0.667 & 8.9$\pm 0.15$ & 8 \\
	4FGL J1635.2+3808 & 248.8168 & 38.1401 & 4C +38.41 & 1.81396 & FSRQ & -0.734 & $9.5\pm0.5$ & 6 \\
	4FGL J1159.5+2914 & 179.884 & 29.2448 & Ton 599 & 0.72474 & FSRQ & -0.932 & $8.5\pm0.5$ & 2 \\
	4FGL J1229.0+0202 & 187.2675 & 2.0454 & 3C 273 & 0.15834 & FSRQ & -0.089 & $8.9\pm0.5$ & 2 \\
	4FGL J1522.1+3144 & 230.5454 & 31.7395 & B2 1520+31 & 1.48875 & FSRQ & -1.134 & 9.4 & 3 \\
	4FGL J1104.4+3812 & 166.1187 & 38.207 & Mkn 421 & 0.03002 & BLL & -3.567 & $8.3\pm0.2$ & 1 \\
	4FGL J0428.6-3756 & 67.173 & -37.9403 & PKS 0426-380 & 1.105 & BLL & -1.673 & 8.6 & 6 \\
	4FGL J0721.9+7120 & 110.4882 & 71.3405 & S5 0716+71 & 0.31 & BLL & $\cdots$ & 8.7 & 10 \\
	4FGL J1555.7+1111 & 238.9313 & 11.1884 & PG 1553+113 & 0.36 & BLL & $\cdots$ & 8.7 & 11 \\
	4FGL J0854.8+2006 & 133.7071 & 20.1159 & OJ 287 & 0.3056 & BLL & -2.472 & $8.8\pm0.5$ & 1 \\
	4FGL J0509.4+0542 & 77.3593 & 5.7041 & TXS 0506+056 & 0.3365 & BLL & $\cdots$ & $8.5\pm0.6$ & 9 
	\enddata 
	\tablecomments{(1)(4) source name, (2)(3) RA and Dec (J2000), (5) redshift, (6) source type, (7) Eddington ratio from \citet{2015MNRAS.450.3568X}, 
(8) black hole masses and their uncertainties (in solar mass) collected from the references in the last column. 
The uncertainty in the relation between stellar velocity dispersion and black hole mass ($M_{\rm BH}-\sigma$) is $\lesssim0.21$ dex \citep{2002ApJ...574..740T}, and we use 0.2 dex. 
The uncertainty on the zero point of the line width-luminosity-mass relation is approximately 0.5 dex \citep{2000ApJ...543L...5G,2001ApJ...555L..79F}. 
The correlation between host galaxy luminosity and black hole mass ($L_{\rm host}-M_{\rm BH}$) has an uncertainty of 0.6 dex \citep{2001MNRAS.327..199M}.
	References: (1) \cite{2004ApJ...615L...9W}, (2) \cite{2006ApJ...637..669L}, (3) \cite{2010MNRAS.402..497G}, (4) \cite{2017MNRAS.469.2305K}, (5) \cite{2018FrASS...5....2S}, (6) \cite{2012MNRAS.421.1764S}, (7)  \cite{2004ApJ...602..103F}, (8) \cite{2012ApJ...748...49S}, (9) \cite{2019MNRAS.484L.104P}, (10) \cite{2018AJ....156...36K}, (11) \cite{2017ApJ...851...33P},(12) \cite{2012NewA...17....8G}.}}
\end{deluxetable*}

\begin{deluxetable*}{cccccccc}
\tiny{
	\caption{Modeling Results for 23 AGNs.\label{tab:modeling results}}
	\tablewidth{20pt}
	\setlength{\tabcolsep}{4mm}{
	\tablehead{
		\colhead{Name} & \colhead{Mean cadence} &\colhead{Model} & \colhead{$\rm AIC_{c}$} &  \multicolumn{3}{c}{Normal Distribution Fitting} \\
		\cmidrule(r){5-7}
		\colhead{} & \colhead{(days)} & \colhead{} & \colhead{} & \colhead{$\mu$} & \colhead{$\sigma$} & \colhead{$\chi^{2}_{\rm red}$}\\
		\colhead{(1)} & \colhead{(2)} & \colhead{(3)} & \colhead{(4)} & \colhead{(5)} & \colhead{(6)} & \colhead{(7)} 
	}
	\startdata
	3C 454.3 & 16.37 & SHO & 2341.31 & $-0.19\pm0.03$ & $0.33\pm0.02$ & 1.904 \\
	& & DRW & 2338.69 & $-0.16\pm0.03$ & $0.41\pm0.02$ & 1.401 \\
	3C 279 & 15.71 & SHO & 2078.40 & $-0.18\pm0.02$ & $0.25\pm0.02$ & 2.320 \\
	& & DRW & 2076.14 & $-0.18\pm0.02$ & $0.25\pm0.02$ & 2.320 \\
	PKS 1424-41 & 16.01 & SHO & 1095.99 & $-0.14\pm0.05$ & $0.57\pm0.04$ & 1.697 \\
	& & DRW & 1093.93 & $-0.14\pm0.05$ & $0.57\pm0.04$ & 1.682 \\
	PKS 1510-089 & 15.66 & SHO & 1859.79 & $-0.18\pm0.03$ & $0.43\pm0.03$ & 1.713 \\
	& & DRW & 1861.70 & $-0.17\pm0.03$ & $0.43\pm0.03$ & 1.731 \\
	PKS 1502+106 & 19.19 & SHO & 873.39 & $-0.11\pm0.06$ & $0.68\pm0.05$ & 1.377 \\
	& & DRW & 877.79 & $-0.10\pm0.05$ & $0.68\pm0.05$ & 1.154 \\
	BL Lac & 16.09 & SHO & 1284.71 & $-0.09\pm0.04$ & $0.57\pm0.03$ & 1.145 \\
	& & DRW & 1282.62 & $-0.08\pm0.04$ & $0.59\pm0.04$ & 1.263 \\
	PKS 2155-304 & 15.35 & SHO & 293.01 & $-0.17\pm0.06$ & $0.74\pm0.05$ & 1.939 \\
	& & DRW & 290.97 & $-0.17\pm0.06$ & $0.74\pm0.05$ & 1.946 \\
	PKS 0537-441 & 16.43 & SHO & 584.69 & $-0.09\pm0.05$ & $0.76\pm0.04$ & 0.939 \\
	& & DRW & 582.62 & $-0.09\pm0.05$ & $0.76\pm0.04$ & 0.975 \\
	3C 66A & 16.09 & SHO & 271.57 & $-0.09\pm0.05$ & $0.83\pm0.05$ & 1.010 \\
	& & DRW & 269.49 & $-0.08\pm0.06$ & $0.84\pm0.05$ & 1.030 \\
	NGC 1275 & 15.20 & SHO & 972.44 & $-0.08\pm0.05$ & $0.78\pm0.04$ & 0.875 \\
	& & DRW & 970.35 & $-0.07\pm0.05$ & $0.79\pm0.04$ & 0.853 \\
	CTA 102 & 18.32 & SHO & 1727.43 & $-0.17\pm0.03$ & $0.33\pm0.02$ & 1.568 \\
	& & DRW & 1725.47 & $-0.22\pm0.03$ & $0.29\pm0.02$ & 2.285 \\
	PKS 0454-234 & 16.15 & SHO & 842.94 & $-0.17\pm0.05$ & $0.74\pm0.05$ & 1.262\\
	& & DRW & 844.64 & $-0.19\pm0.06$ & $0.69\pm0.05$ & 1.582\\
	4C +21.35 & 24.13 & SHO & 993.40 & $-0.20\pm0.03$ & $0.35\pm0.03$ & 1.402 \\
	& & DRW & 991.29 & $-0.20\pm0.03$ & $0.35\pm0.03$ & 1.244\\
	4C +38.41 & 17.22 & SHO & 991.39 & $-0.13\pm0.04$ & $0.51\pm0.04$ & 1.877\\
	& & DRW & 989.32 & $-0.13\pm0.04$ & $0.49\pm0.04$ & 1.961\\
	Ton 599 & 22.04 & SHO & 733.25 & $-0.14\pm0.05$ & $0.62\pm0.05$ & 1.386\\
	& & DRW & 734.89 & $-0.13\pm0.05$ & $0.58\pm0.04$ & 1.418\\
	3C 273 & 22.35 & SHO & 1083.29 & $-0.20\pm0.03$ & $0.38\pm0.03$ & 1.143 \\
	& & DRW & 1081.02 & $-0.20\pm0.03$ & $0.37\pm0.02$ & 1.026 \\
	B2 1520+31 & 22.15 & SHO & 605.98 & $-0.19\pm0.05$ & $0.63\pm0.04$ & 1.110 \\
	& & DRW & 603.86 & $-0.16\pm0.06$ & $0.69\pm0.05$ & 1.097 \\
	Mkn 421 & 15.15 & SHO & 508.63 & $-0.10\pm0.05$ & $0.85\pm0.04$ & 0.791 \\
	& & DRW & 508.32 & $-0.09\pm0.05$ & $0.84\pm0.04$ & 0.753 \\
	PKS 0426-380 & 15.76 & SHO & 710.85 & $-0.15\pm0.06$ & $0.77\pm0.05$ & 1.471\\
	& & DRW & 709.48 & $-0.13\pm0.05$ & $0.76\pm0.04$ & 1.093\\
	S5 0716+71 & 16.37 & SHO & 758.54 & $-0.18\pm0.08$ & $0.85\pm0.07$ & 1.630 \\
	& & DRW & 756.39 & $-0.19\pm0.08$ & $0.85\pm0.07$ & 1.612 \\
	PG 1553+113 & 15.15 & SHO & -351.89 & $-0.02\pm0.05$ & $0.99\pm0.05$ & 0.602 \\
	& & DRW & -353.93 & $-0.03\pm0.06$ & $0.97\pm0.05$ & 0.71\\
	OJ 287 & 26.34 & SHO & 311.75 & $-0.30\pm0.05$ & $0.54\pm0.05$ & 1.475 \\
	& & DRW & 310.54 & $-0.30\pm0.05$ & $0.52\pm0.04$ & 1.662\\
	TXS 0506+056 & 23.26 & SHO & 246.85 & $-0.12\pm0.08$ & $0.77\pm0.07 $& 1.891 \\
	& & DRW & 244.76 & $-0.11\pm0.08$ & $0.78\pm0.07$ & 1.892
    \enddata
	\tablecomments{ (1) source name, (2) the mean cadence of LC, (3) model, (4) AIC$_{\rm c}$, 
	(5)(6)(7) the results and the reduced $\chi^{2}$ of normal distribution fitting to the standard residuals. 
		}}}
\end{deluxetable*}

\begin{deluxetable*}{cccccccc}
	\tablecaption{Posterior Parameters for 23 AGNs.\label{tab:Posterior Parameters}}
	\tablewidth{30pt}
	\setlength{\tabcolsep}{2mm}{
	\tablehead{
		\colhead{Name} &\colhead{Model} & \multicolumn{3}{c}{Parameter of SHO} &  \multicolumn{2}{c}{Parameter of DRW} &\colhead{Damping timescale}  \\
		\cmidrule(r){3-5}\cmidrule(r){6-7}
		\colhead{} & \colhead{} & \colhead{ln $S_{0}$} & \colhead{ln Q} & \colhead{ln $\omega_{0}$} & \colhead{ln $\sigma_{\rm DRW}$} & \colhead{ln $\tau_{\rm DRW}$} & \colhead{(uncorrected)} \\
		\colhead{(1)} & \colhead{(2)} & \colhead{(3)} & \colhead{(4)} & \colhead{(5)} & \colhead{(6)} & \colhead{(7)} & \colhead{(8)}
	}
	\startdata
	3C 454.3 & SHO & $10.60^{+0.41}_{-0.34}$ & $-1.75^{+0.41}_{-1.08}$ & $-2.44^{+0.97}_{-0.27}$ & $\cdots$ & $\cdots$ & $\cdots$ \\
	& DRW & $\cdots$ & $\cdots$ & $\cdots$ & $2.87^{+0.10}_{-0.09}$ & $4.30^{+0.22}_{-0.19}$ & $73.70^{+16.21}_{-14.00}$ \\
	3C 279 & SHO & $7.45^{+0.23}_{-0.21}$ & $-2.07^{+0.47}_{-0.36}$ & $-0.98^{+0.34}_{-0.45}$ & $\cdots$ & $\cdots$ & $\cdots$ \\
	& DRW & $\cdots$ & $\cdots$ & $\cdots$ & $1.85^{+0.05}_{-0.05}$ & $3.06^{+0.15}_{-0.14}$ & $21.33^{+3.20}_{-2.99}$ \\
	PKS 1424-41 & SHO & $7.28^{+0.58}_{-0.45}$ & $-3.28^{+0.79}_{-0.69}$ & $-1.47^{+0.66}_{-0.76}$ & $\cdots$ & $\cdots$ & $\cdots$ \\
	& DRW & $\cdots$ & $\cdots$ & $\cdots$ & $0.93^{+0.14}_{-0.11}$ & $4.72^{+0.30}_{-0.24}$ & $112.17^{+33.65}_{-26.92}$  \\
	PKS 1510-089 & SHO &$7.57^{+0.31}_{-0.27}$ & $-1.46^{+0.42}_{-0.96}$ & $-2.14^{+0.88}_{-0.29}$ & $\cdots$ & $\cdots$ & $\cdots$ \\
	& DRW & $\cdots$ & $\cdots$ & $\cdots$ & $1.65^{+0.07}_{-0.06}$ & $3.68^{+0.17}_{-0.16}$ & $39.65^{+6.74}_{-6.34}$ \\
	PKS 1502+106 & SHO & $7.91^{+1.08}_{-0.68}$ & $-3.89^{+0.85}_{-0.79}$ & $-1.54^{+0.71}_{-0.82}$ & $\cdots$ & $\cdots$ & $\cdots$ \\
	& DRW & $\cdots$ & $\cdots$ & $\cdots$ & $0.91^{+0.24}_{-0.16}$ & $5.33^{+0.50}_{-0.34}$ & $206.44^{+103.22}_{-70.19}$ \\
	BL Lac & SHO & $6.74^{+0.44}_{-0.36}$ & $-3.22^{+0.82}_{-0.71}$ & $-1.02^{+0.70}_{-0.79}$ & $\cdots$ & $\cdots$ & $\cdots$ \\
	& DRW & $\cdots$ & $\cdots$ & $\cdots$ & $0.91^{+0.10}_{-0.09}$ & $4.23^{+0.24}_{-0.20}$ & $68.72^{+16.49}_{-13.74}$ \\
	PKS 2155-304 & SHO & $2.66^{+0.41}_{-0.34}$ & $-2.58^{+0.77}_{-0.73}$ & $-1.55^{+0.71}_{-0.72}$ & $\cdots$ & $\cdots$ & $\cdots$ \\
	& DRW & $\cdots$ & $\cdots$ & $\cdots$ & $-1.07^{+0.10}_{-0.08}$ & $4.12^{+0.24}_{-0.20}$ & $61.56^{+14.77}_{-12.31}$ \\
	PKS 0537-441 & SHO & $6.35^{+0.93}_{-0.61}$ & $-4.01^{+0.82}_{-0.73}$ & $-1.48^{+0.67}_{-0.80}$ & $\cdots$ & $\cdots$ & $\cdots$ \\
	& DRW & $\cdots$ & $\cdots$ & $\cdots$ & $0.09^{+0.21}_{-0.14}$ & $5.41^{+0.44}_{-0.31}$ & $223.63^{+98.40}_{-69.33}$ \\
	3C 66A & SHO & $3.15^{+0.50}_{-0.40}$ & $-3.11^{+0.75}_{-0.64}$ & $-1.40^{+0.61}_{-0.72}$ & $\cdots$ & $\cdots$ & $\cdots$ \\
	& DRW & $\cdots$ & $\cdots$ & $\cdots$ & $-1.02^{+0.12}_{-0.10}$ & $4.48^{+0.28}_{-0.24}$ & $88.23^{+24.71}_{-21.18}$ \\
	NGC 1275 & SHO & $5.51^{+0.42}_{-0.35}$ & $-3.00^{+069}_{-0.56}$ & $-1.28^{+0.54}_{-0.66}$ & $\cdots$ & $\cdots$ & $\cdots$ \\
	& DRW & $\cdots$ & $\cdots$ & $\cdots$ & $0.27^{+0.10}_{-0.08}$ & $4.27^{+0.23}_{-0.19}$ & $71.52^{+16.45}_{-13.59}$ \\
	CTA 102 & SHO & $9.31^{+0.43}_{-0.35}$ & $-2.79^{+0.76}_{-0.73}$ & $-1.56^{+0.71}_{-0.70}$ & $\cdots$ & $\cdots$ & $\cdots$ \\
	& DRW & $\cdots$ & $\cdots$ & $\cdots$ & $2.15^{+0.11}_{-0.09}$ & $4.34^{+0.23}_{-0.20}$ & $76.71^{+17.64}_{-15.34}$ \\
	PKS 0454-234 & SHO & $5.28^{+0.43}_{-0.35}$ & $-1.84^{+0.48}_{-1.17}$ & $-2.44^{+1.08}_{-0.34}$ & $\cdots$ & $\cdots$ & $\cdots$ \\
	& DRW & $\cdots$ & $\cdots$ & $\cdots$ & $0.17^{+0.11}_{-0.09}$ & $4.36^{+0.24}_{-0.20}$ & $78.26^{+18.78}_{-15.65}$ \\
	4C +21.35 & SHO & $7.41^{+0.43}_{-0.36}$ & $-2.71^{+0.71}_{-0.59}$ & $-1.32^{+0.57}_{-0.67}$ & $\cdots$ & $\cdots$ & $\cdots$ \\
	& DRW & $\cdots$ & $\cdots$ & $\cdots$ & $1.35^{+0.10}_{-0.09}$ & $4.02^{+0.23}_{-0.20}$ & $55.70^{+12.81}_{-11.14}$ \\
	4C +38.41 & SHO & $6.57^{+0.52}_{-0.41}$ & $-2.95^{+0.59}_{-0.50}$ & $-1.66^{+0.46}_{-0.55}$ & $\cdots$ & $\cdots$  $\cdots$ \\
	& DRW & $\cdots$ & $\cdots$ & $\cdots$ & $0.63^{+0.12}_{-0.10}$ & $4.60^{+0.27}_{-0.22}$ & $99.48^{+26.86}_{-21.89}$ \\
	Ton 599 & SHO & $5.72^{+0.47}_{-0.39}$ & $-1.84^{+0.52}_{-1.17}$ & $-.2.39^{+1.09}_{-0.35}$ & $\cdots$ & $\cdots$  & $\cdots$ \\
	& DRW & $\cdots$ & $\cdots$ & $\cdots$ & $0.42^{+0.12}_{-0.10}$ & $4.30^{+0.26}_{-0.22}$ & $73.70^{+19.16}_{-16.21}$\\
	3C 273 & SHO & $6.14^{+0.30}_{-0.27}$ & $-2.43^{+0.49}_{-0.39}$ & $-1.00^{+0.35}_{-0.47}$ & $\cdots$ & $\cdots$ & $\cdots$ \\
	& DRW & $\cdots$ & $\cdots$ & $\cdots$ & $1.00^{+0.07}_{-0.06}$ & $3.44^{+0.18}_{-0.17}$ & $31.19^{+5.61}_{-5.30}$ \\
	B2 1520+31 & SHO & $5.19^{+0.46}_{-0.38}$ & $-2.95^{+0.67}_{-0.56}$ & $-1.27^{+0.53}_{-0.64}$ & $\cdots$ & $\cdots$ & $\cdots$ \\
	& DRW & $\cdots$ & $\cdots$ & $\cdots$ & $0.13^{+0.11}_{-0.09}$ & $4.22^{+0.26}_{-0.22}$ & $68.03^{+17.69}_{-14.97}$ \\
	Mkn 421 & SHO & $3.20^{+0.34}_{-0.29}$ & $-1.98^{+0.70}_{-0.93}$ & $-1.88^{+0.90}_{-0.56}$ & $\cdots$ & $\cdots$ & $\cdots$ \\
	& DRW & $\cdots$ & $\cdots$ & $\cdots$ & $-0.66^{+0.08}_{-0.07}$ & $3.89^{+0.19}_{-0.17}$ & $48.91^{+9.29}_{-8.80}$ \\
	PKS 0426-380 & SHO & $5.25^{+0.51}_{-0.41}$ & $-2.76^{+0.86}_{-0.95}$ & $-1.89^{+0.93}_{-0.77}$ & $\cdots$ & $\cdots$ & $\cdots$ \\
	& DRW & $\cdots$ & $\cdots$ & $\cdots$ & $-0.03^{+0.12}_{-0.10}$ & $4.65^{+0.27}_{-0.22}$ & $104.58^{+28.24}_{-23.01}$ \\
	S5 0716+71 & SHO & $3.23^{+0.25}_{-0.23}$ & $-2.08^{+0.53}_{-0.43}$ & $-1.09^{+0.41}_{-0.51}$ & $\cdots$ & $\cdots$ & $\cdots$ \\
	& DRW & $\cdots$ & $\cdots$ & $\cdots$ & $-0.32^{+0.06}_{-0.05}$ & $3.18^{+0.16}_{-0.15}$ & $24.05^{+3.85}_{-3.61}$ \\
	PG 1553+113 & SHO & $2.02^{+1.35}_{-0.76}$ & $-3.49^{+1.27}_{-1.22}$ & $-2.19^{+1.15}_{-1.20}$ & $\cdots$ & $\cdots$ & $\cdots$ \\
	& DRW & $\cdots$ & $\cdots$ & $\cdots$ & $-2.12^{+0.27}_{-0.17}$ & $5.51^{+0.65}_{-0.43}$ & $247.15^{+160.65}_{-106.27}$ \\
	OJ 287 & SHO & $2.80^{+0.36}_{-0.32}$ & $-1.96^{+0.77}_{-0.86}$ & $-1.76^{+0.84}_{-0.67}$ & $\cdots$ & $\cdots$ & $\cdots$ \\
	& DRW & $\cdots$ & $\cdots$ & $\cdots$ & $-0.79^{+0.09}_{-0.08}$ & $3.73^{+0.23}_{-0.21}$ & $41.68^{+9.59}_{-8.75}$ \\
	TXS 0506+056 & SHO & $3.44^{+0.62}_{-0.48}$ & $-3.29^{+0.80}_{-0.70}$ & $-1.45^{+0.66}_{-0.78}$ & $\cdots$ & $\cdots$ & $\cdots$ \\
	& DRW & $\cdots$ & $\cdots$ & $\cdots$ & $-0.99^{+0.14}_{-0.11}$ & $4.70^{+0.37}_{-0.31}$ & $109.95^{+40.68}_{-34.08}$ 
	\enddata
	\tablecomments{ 
	(1) source name, (2) model, (3)(4)(5) posterior parameters of modeling LCs with SHO model, (6)(7) posterior parameters of modeling LCs with DRW model, (8) damping timescale. The uncertainties of model parameters and damping timescales represent $1\sigma$ confidence intervals.  
		}}
\end{deluxetable*}

\section{Fermi-LAT Data Analysis} \label{sec:style}

We collect 23 bright $\gamma$-ray AGNs with significant variability in the LAT data. 
The information of these sources are listed in the Table \ref{tab:information}. 

All data analyzed here come from the LAT 8 yr Source Catalog (4FGL;  \citealt{2020ApJS..247...33A}), spanning the time range of MJD 54682-59332 which gives in total $\sim$ 12.7 yr of data in the energy range of 0.1-500 GeV. 
We consider only SOURCE class events (evclass=128) and event type three (evtype=3) from the region of interest (ROI) at 15$^{\circ}$ for each source.
The maximum zenith angle is set to be $90^{\circ}$ to avoid contaminating from Earth's limb.
DATAQUAL $\textgreater$0 and LATCONFIG == 1 options are chosen to ensure the good data quality and the proper time intervals. 
The instrument response function P8R3$\_$SOURCE$\_$V3 is applied in the analysis. 
We use the Galactic
(gll$\_$iem$\_$v07.fits) and the extragalactic (iso$\_$P8R3$\_$SOURCE$\_$V3$\_$v1.txt) diffuse
emission models, which are the latest Pass 8 background models.
We use the binned maximum likelihood analysis\footnote{https://fermi.gsfc.nasa.gov/ssc/data/analysis/scitools/binned\_likelihood\_tutorial.html} \citep{2009ApJS..183...46A}, 
which is the preferred method for most types of LAT analysis.
It is a three-dimensional
maximum likelihood algorithm, i.e., events are binned into channels of energy and position in sky \citep{2009ApJS..183...46A}, 
and a maximum likelihood optimization technique is performed to determine the best-fit parameters and the Test Statistic $\rm TS=2\Delta log(likelihood)$  between models with and without the source \citep{1996ApJ...461..396M}.
In the fitting, we use the spectral model of the LogParabola (LP) form ($dN/dE=N_{0}\left(E/E_{\rm b}\right)^{-\alpha + \beta \log(E / E_{\rm b})}$).

\section{Stochastic Process Method} \label{sec:model}

We use the stochastic process method implemented in {\it celerite} package \citep{2017AJ....154..220F}.
In {\it celerite} package, a specific and stationary kernel function (i.e., covariance function) is required, which can be defined by users.

\subsection{DRW Model} \label{subsec:DRW}
The DRW process is described by a first order stochastic differential equation \citep[see details in][]{2009ApJ...698..895K,2019PASP..131f3001M}.
It represents a competition between a process trying to maintain an equilibrium state and a perturbation making the system out of stability.
It is sometimes also written as a Langevin equation of the form
\begin{equation}\label{eqDRW}
\left[\frac{d}{dt}+\frac{1}{\tau_{\rm DRW}}\right]y(t)=\sigma_{\rm DRW}\epsilon(t)\;,
\end{equation}
where $\tau_{\rm DRW}$ is the damping timescale and $\sigma_{\rm DRW}$ is the amplitude of random perturbations.

Following the setting in {\it celerite} package, 
the covariance function for the DRW is written as
\begin{equation}\label{eq4}
k(t_{nm})=a\cdot \exp(-t_{nm}c)\;,
\end{equation}
where $t_{nm} = |t_{n}-t_{m}|$ is the time lag between measurements $m$ and $n$, $a=2\sigma_{\rm DRW}^{2}$, 
and $c=1/\tau_{\rm DRW}$.
The PSD is written as \citep{2017AJ....154..220F}
\begin{equation}\label{eq5}
S(\omega)=\sqrt{\frac{2}{\pi}}\frac{a}{c}\frac{1}{1+(\omega/c)^2}\ .
\end{equation}
The DRW PSD is a broken power-law form, and the index changes from 0 at low frequencies  to -2 at high frequencies.
The broken frequency $f_{\rm b}$ corresponds to the damping timescale, $\tau_{\rm DRW}$ = 1/(2$\pi f_{\rm b}$).
  
\subsection{SHO Model} \label{subsec:SHO}

The dynamics of a stochastically driven damped simple harmonic oscillator (SHO) provides a physically motivated model, as it can describe the variability driven by
noisy physical processes, which grows most strongly at the
characteristic timescale but is also damped owing to dissipation
in the system \citep{2017AJ....154..220F}.
The differential equation for this system is 
\begin{equation}\label{eq2}
\left[\frac{d^{2}}{dt^{2}}+\frac{\omega_{0}}{Q}\frac{d}{dt}+\omega_{0}^{2}\right]y(t)=\epsilon(t)\;,
\end{equation}
with the frequency of the undamped oscillator $\omega_{0}$, the quality factor of the oscillator $Q$, and a stochastic driving force $\epsilon(t)$. 
When the $\epsilon(t)$ is white noise, the PSD of this process is 
\begin{equation}\label{eq3}
S(\omega)=\sqrt{\frac{2}{\pi}}\frac{S_{0}\omega_{0}^{4}}{(\omega^{2}-\omega_{0}^{2})^{2}+\omega^{2}\omega_{0}^{2}/Q^{2}}\;,
\end{equation}
where $S_{0}$ is proportional to the power at $\omega=\omega_{0}$.
The SHO PSD is complex.
For the over-damped SHO ($Q<0.5$), it is also a broken power-law form at low frequencies, very similar to the DRW PSD, while
at high frequencies, the index can be as small as $\sim-4$ \citep{2017MNRAS.470.3027K,2019PASP..131f3001M}.
For the under-damped SHO ($Q>0.5$), a Lorentzian appears in the PSD, i.e., a QPO signal  \citep{2017AJ....154..220F,2019PASP..131f3001M}.

\subsection{Model Selection} \label{subsec:AIC}
Akaike information criterion (AIC) estimates the relevant information that is lost when models are used to represent the underlying processes that generate the data. 
It is an estimator of the relative quality of models for a given set of data: the less information a model loses, the higher the quality of that model.
We use the corrected AIC (AIC$_{\rm c}$) to perform model selection, which is given by
\begin{equation}\label{eq6}
{\rm AIC}_{\rm c}=2k-2\log L+\frac{2k(k+1)}{n-k-1}\;,
\end{equation}
where $k$ is the number of model parameters, $L$ is the maximum likelihood, and $n$ is the number of data points. 
A preferred model is one that minimizes AIC$_{\rm c}$. 
It is accepted that $\Delta$(AIC$_{\rm c}$)$\gtrsim$10 is a difference substantial enough to prefer the model with smaller AIC$_{\rm c}$ \citep{doi:10.1177/0049124104268644,2014ApJ...786..143S}.

\section{Results} \label{sec:results}

In the fittings, Markov chain Monte Carlo (MCMC) implemented in the package emcee \footnote{\url{https://github.com/dfm/emcee}} \citep{2013PASP..125..306F} is used to sample the posterior probability density in our analysis. 
The priors for the parameters are assumed to be flat.
We run the MCMC sampler for 50,000 steps of which the first 20,000 steps are taken as burn-in.
We calculate the maximum likelihood for optimization which is executed 100 times to resolve possible instability of the algorithm L-BFGS-B, and then calculate AIC$_{\rm c}$ by using the maximum value among the 100 values. 

For the data from MJD 54682 to 59332, the 15-day binning LCs of the 23 AGNs are structured by performing the binned likelihood method for each time bin. 
The time bins having TS value of $\geq $ 25 are selected here, in order to get reliable and high signal-to-noise ratio results \citep[e.g.,][]{2020ApJS..247...27K}.
In Table \ref{tab:modeling results}, we report the mean cadence for each LC.

\begin{figure}
    \centering
    {\includegraphics[width=8cm]{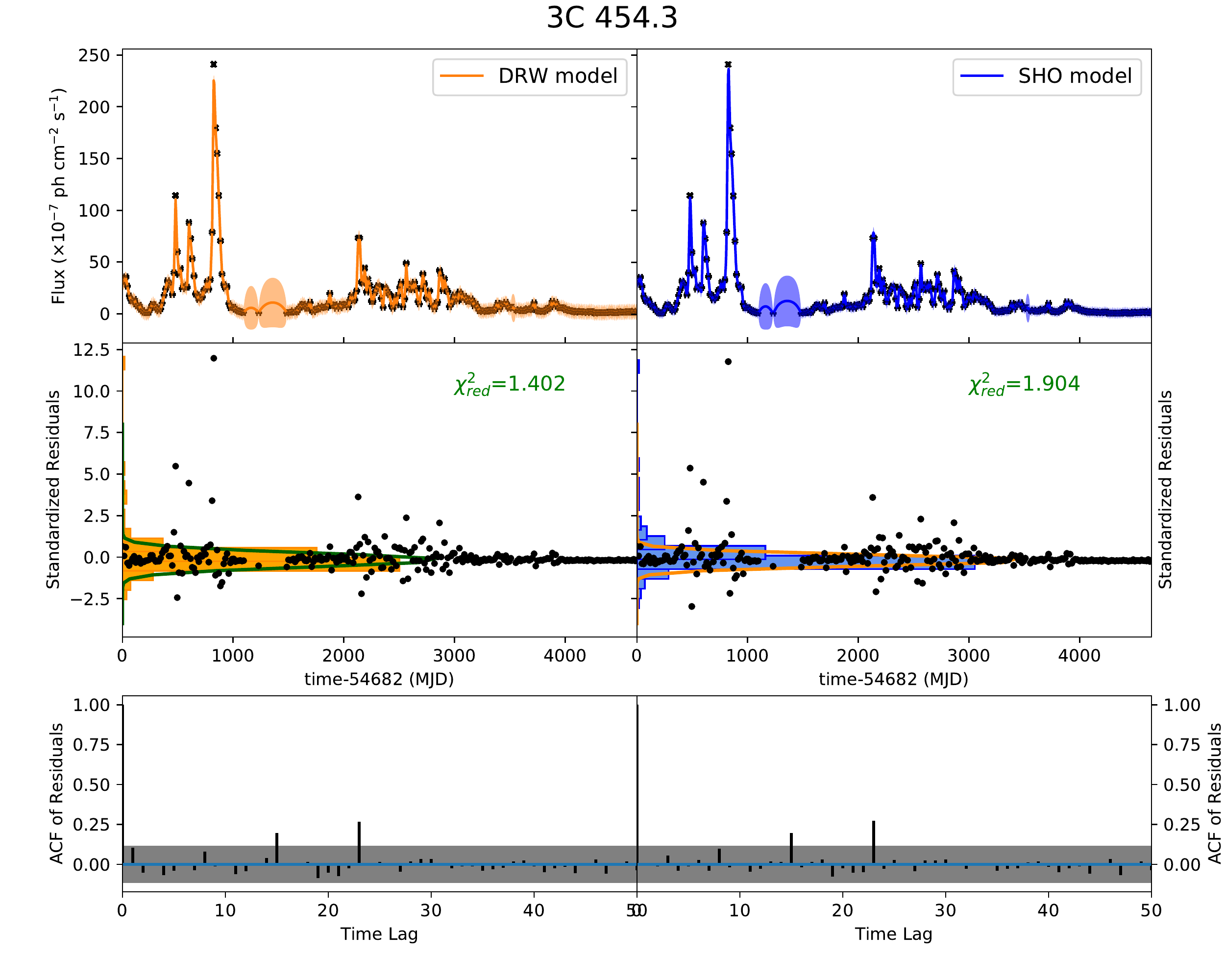}}
    {\includegraphics[width=8cm]{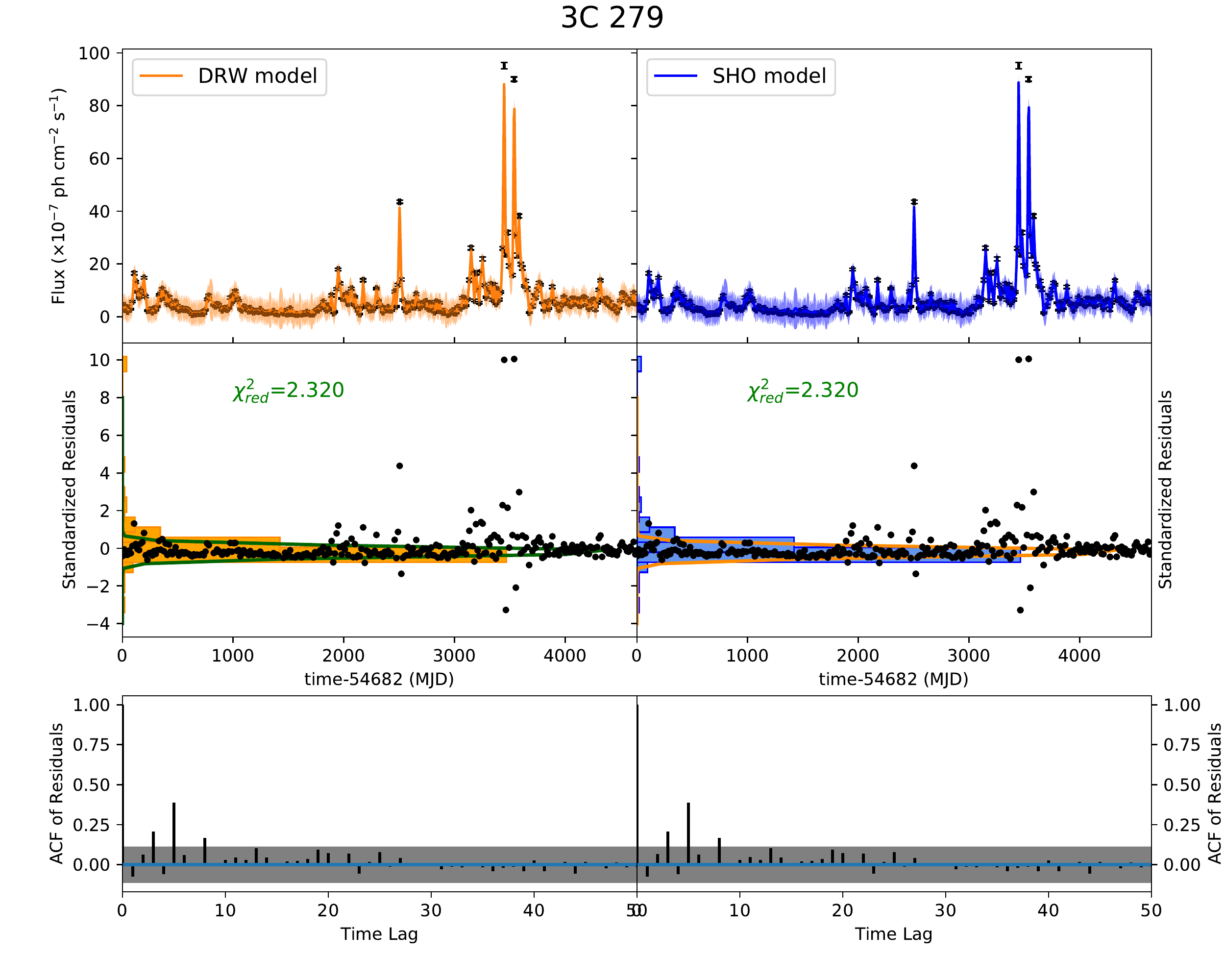}}
    \caption{Fitting results of 3C 454.3 and 3C 279 for example. 
    For each source, the left column presents results obtained from DRW model and we give the LAT LC (black points) and the modeled LC (orange line) in the top panel. 
    In the middle panel, we show the standardized residuals (black points), the probability density of standardized residuals (orange histogram) as well as the best-fit normal distribution (green solid line). The reduced $\chi^{2}$ is labeled in the figure. 
    The ACF of residuals with the 95$\%$ confidence limits of the white noise (the gray region) are shown in the bottom panel. 
    The results obtained by the SHO model are given in the right column. The symbols and lines are the same as the left column but with different colors.
\label{fig:celerite fit}}
\end{figure}

Each LC is fitted with the SHO and DRW models, respectively.
The goodness of the fit is assessed by analyzing the probability densities of the standardized residuals and the auto-correlation function (ACF) of the standardized residuals.
The distribution of the standardized residuals is fitted by a normal distribution. 
The parameters and the reduced $\chi^{2}$ ($\chi^{2}_{\rm red}$) are given in Table \ref{tab:modeling results}. 
It is shown that the distribution of the standardized residuals is in good agreement with the normal distribution with the mean value close to zero and the standard deviation less than one.
In Figure~\ref{fig:celerite fit}, we show the fitting results for 3C 454.3 and 3C 279 for an example.
The ACFs of the residuals are randomly distributed around zero, and are almost inside the 95$\%$ confidence limits of the white noise.
It indicates that the model has captured the characteristics of the time series.
 It is notable that the standardized residuals corresponding to the two or three highest flux are large.
 The DRW and SHO models likely fail to describe the brightest flares.
 When we fit the LC excluded the two or three highest flux points, 
 the modeling results are unchanged. 

\begin{figure}
    \centering
    {\includegraphics[width=8cm]{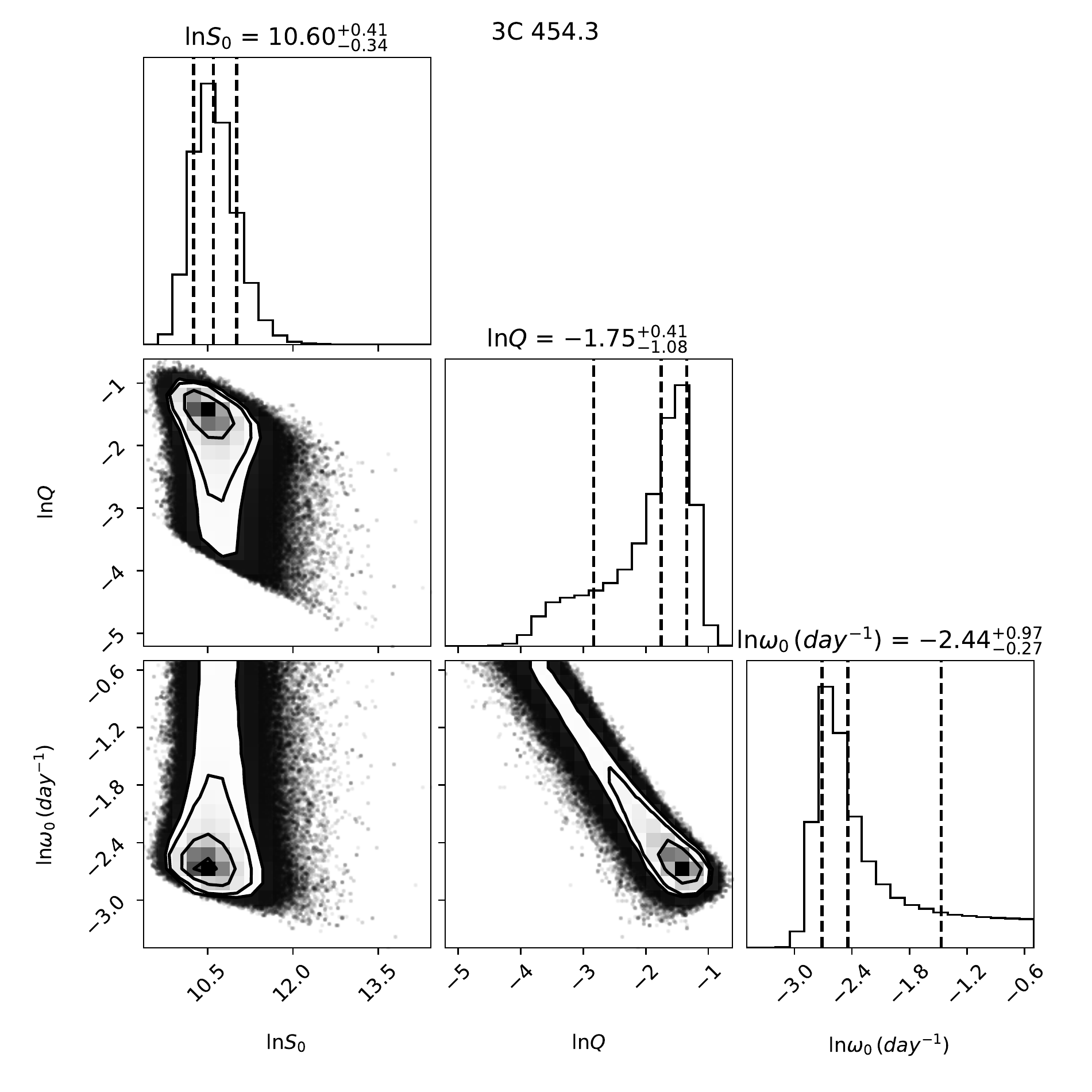}} 
    {\includegraphics[width=8cm]{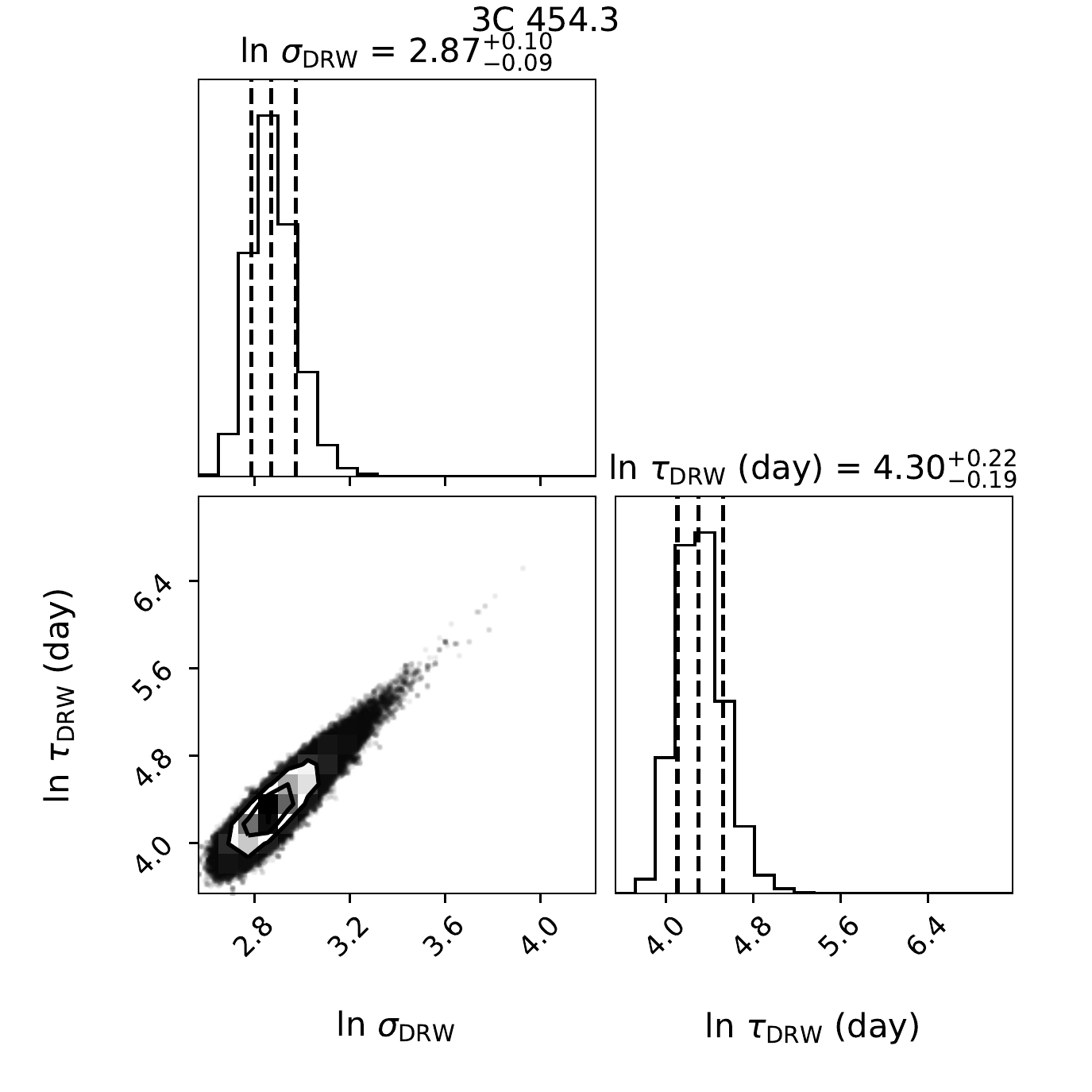}}
    {\includegraphics[width=8cm]{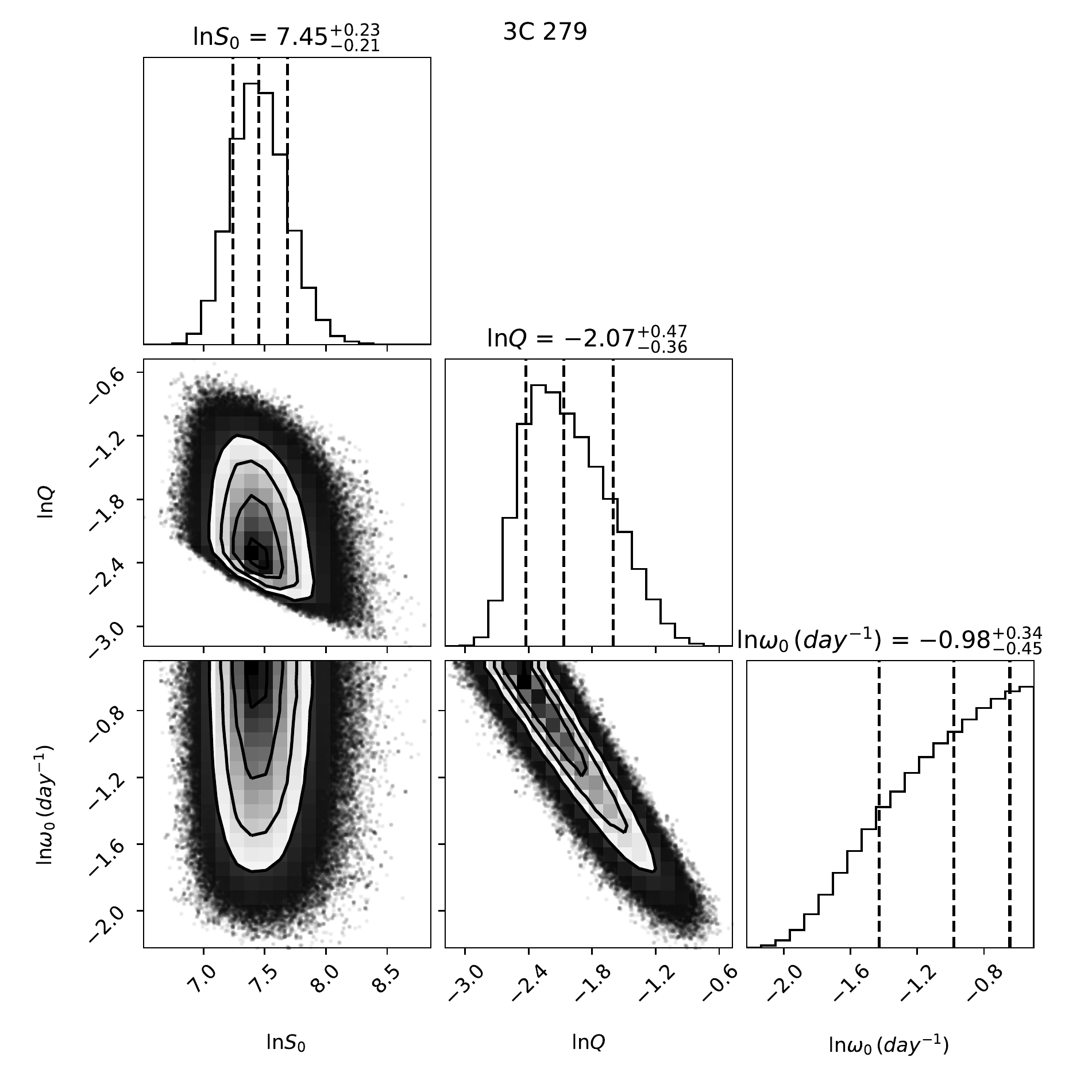}}
    {\includegraphics[width=8cm]{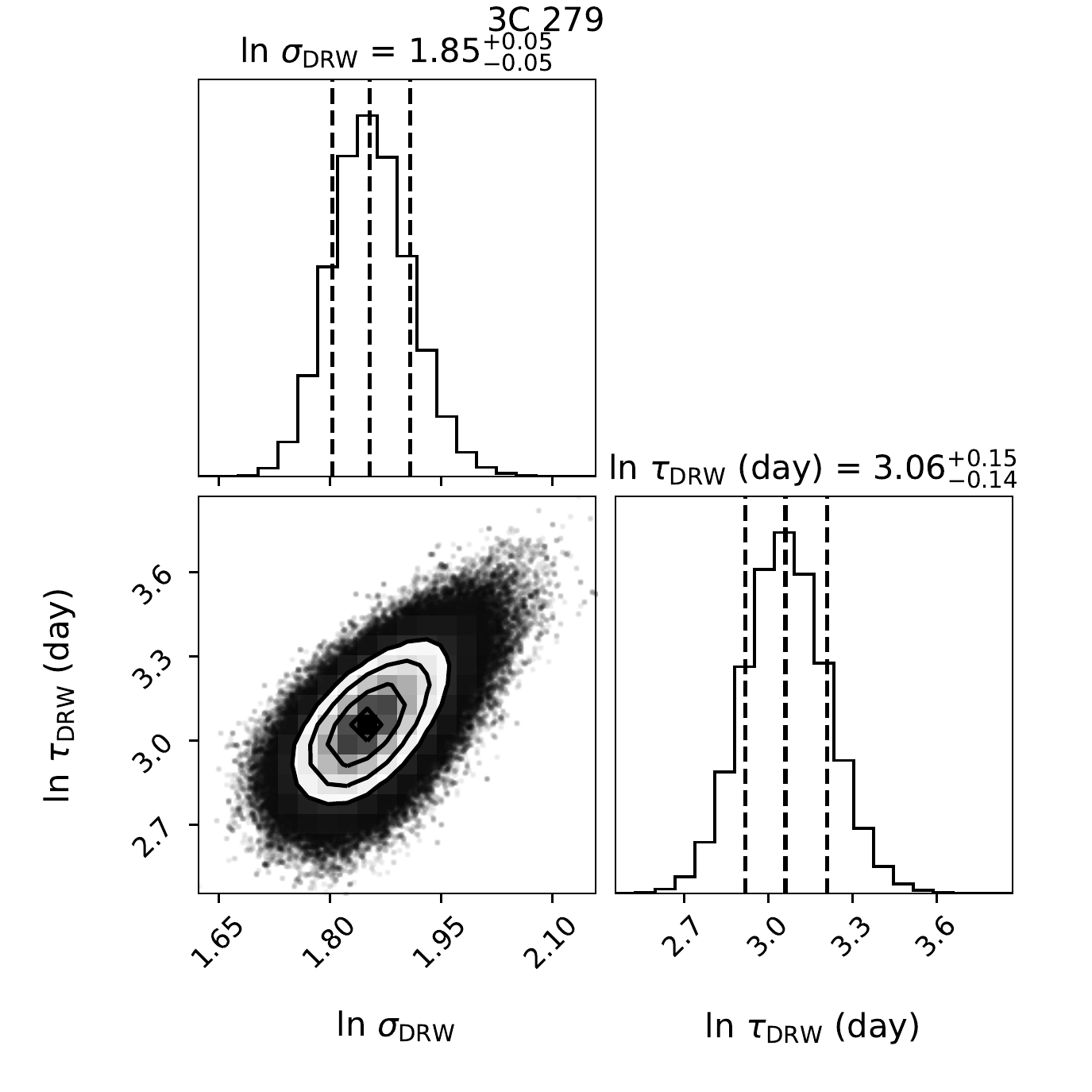}}
\caption{Posterior probability densities of model parameters for SHO (left) and DRW (right). The top panels are for 3C 454.3, and the bottom panels are for 3C 279.
             The vertical dotted lines mark the median value and 68$\%$ confidence intervals of the parameters distribution. 
\label{fig:SHOparam distribution}}
\end{figure}

We give the posterior probability density distribution of parameters resulting from the SHO and DRW modelings in Figure \ref{fig:SHOparam distribution}. 
The values of the model parameters are given in Table \ref{tab:Posterior Parameters}. 
One can see that the model parameters in DRW model are constrained well, although the two parameters are degenerate.
While in the SHO model, the strong degeneracy between $\omega_{0}$ and $Q$ leads to large uncertainties on the two parameters.
In some cases, the large uncertainties on $Q$ cause the unilateral distribution of $\omega_{0}$, and the upper limit for $\omega_{0}$  is meaningless. 
For all sources, the SHO model is constrained in the over-damped mode ($Q<0.5$).

\begin{figure}
    \centering
       {\includegraphics[width=8.5cm]{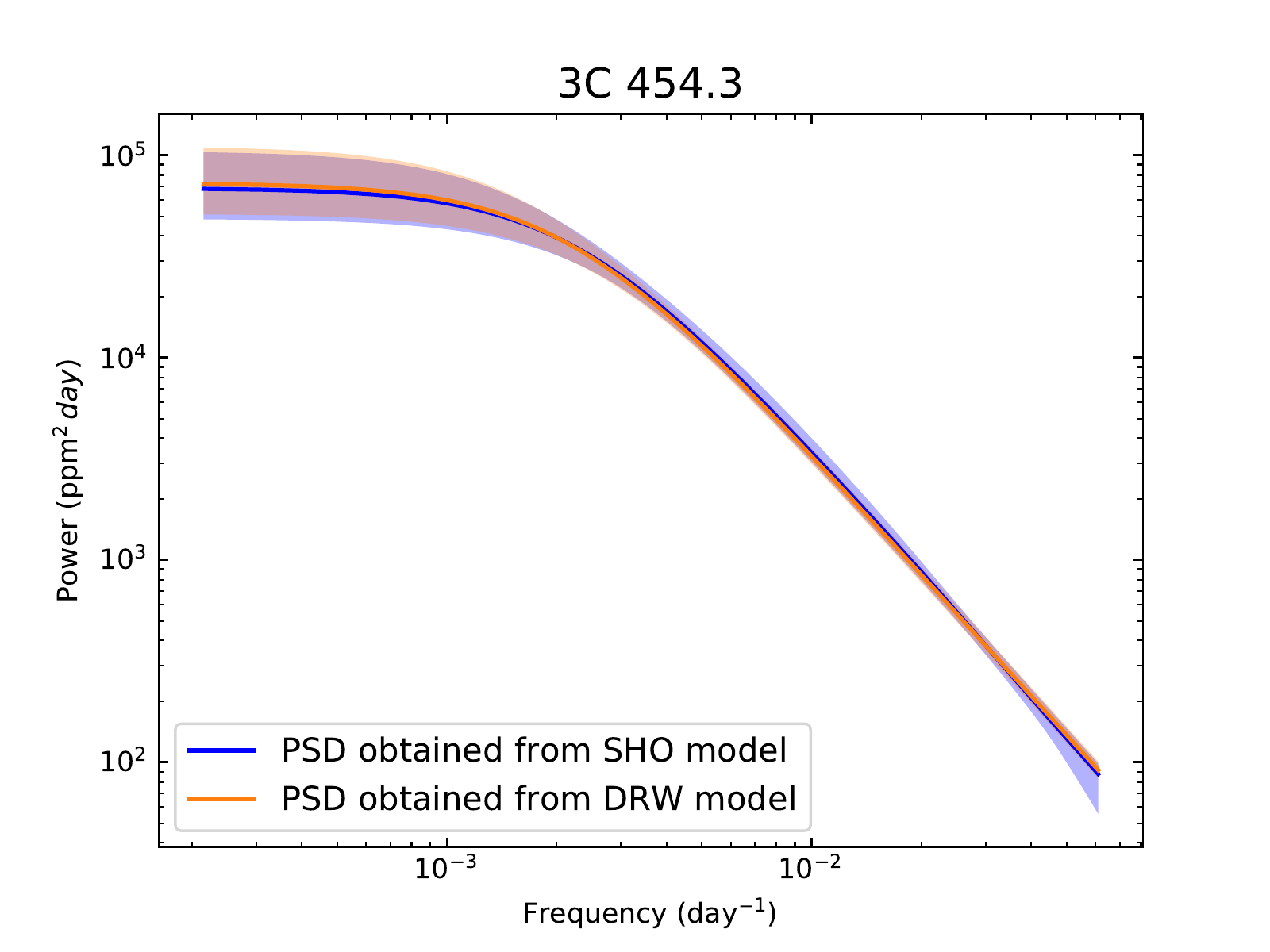}}
       {\includegraphics[width=8cm]{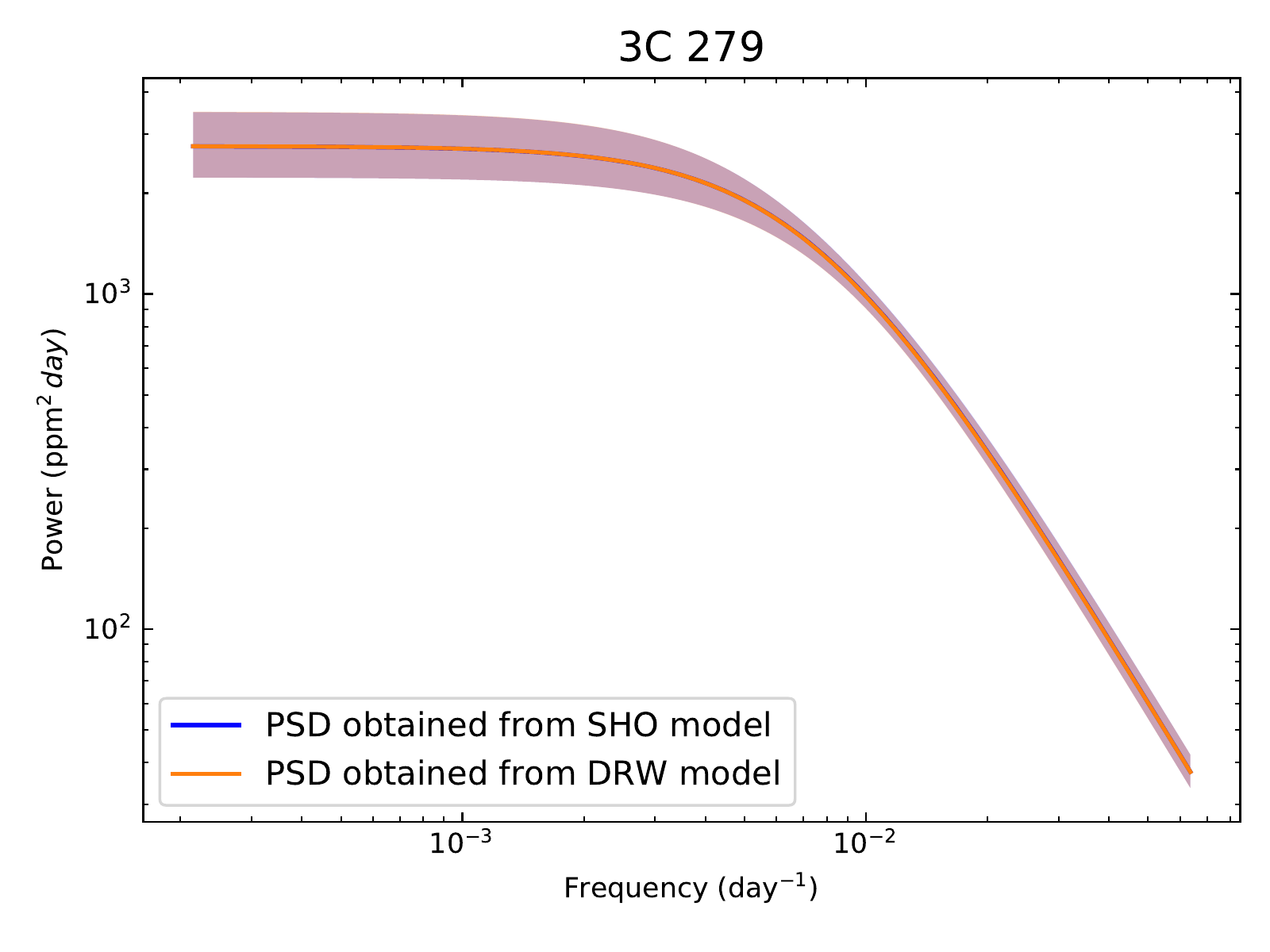}}
  \caption{PSDs of the $\gamma$-ray LCs of 3C 454.3 (left) and 3C 279 (right) constructed from the modeling results with SHO (blue) and DRW (orange). The corresponding color region denotes $1\sigma$ confidence interval.
    \label{fig:psd3C454.3}}
\end{figure}

In Figure~\ref{fig:psd3C454.3}, we show the PSDs for 3C 454.3 and 3C 279 constructed from the modeling results with SHO and DRW models.
The PSDs for SHO and DRW are almost the same. The index changes from 0 at frequencies below $f_{\rm b}$ to -2 at frequencies above $f_{\rm b}$.

The difference between the $\rm AIC_c$ for SHO and DRW are small (Table \ref{tab:modeling results}), indicating that the fittings of the two models are comparable.
However, the poor constrained $\omega_{0}$ and $Q$ suggest that the DRW model is preferred over the SHO model.

\begin{figure}
    \centering
       {\includegraphics[width=12cm]{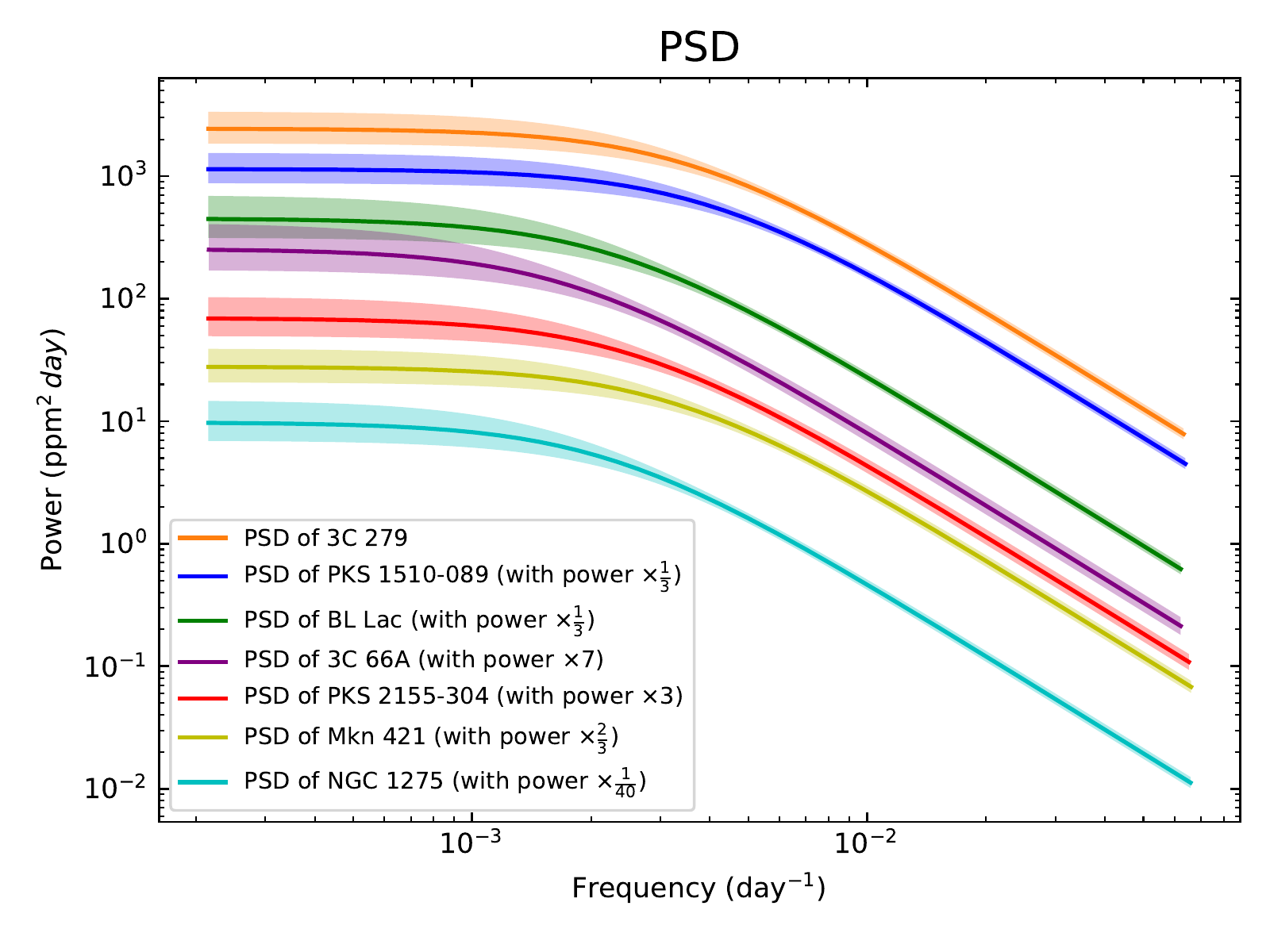}} 
  \caption{DRW PSDs for different types of AGNs.
    \label{fig:psd3}}
\end{figure}

There is no significant difference among the forms of the PSDs from different types of AGNs (Figure~\ref{fig:psd3}).
The PSDs for the 23 sources are typical DRW PSD, and the broken frequencies are between $0.001\ \rm day^{-1}$ and $0.01\ \rm day^{-1}$.

\section{Variability Characteristic Timescales in Relativistic Jets} \label{sec:tau-mass fit} 

Characteristic timescale is an important parameter in source variability.
Our results show that DRW model can describe the $\gamma$-ray variability of 23 targets successfully, and the higher-order model SHO is unnecessary. 
From the fittings with the DRW model, we can obtain the characteristic timescales for the 23 sources.
The characteristic (damping) timescales with errors in $1\sigma$ are given in Table \ref{tab:Posterior Parameters}. 
One can see that the timescales are between 20 days and 250 days.
Note that the measurement of the damping timescale can be biased by the insufficient length of the LC \citep{2017A&A...597A.128K,2021ApJ...907...96S}.
Moreover, \cite{2021Sci...373..789B} found that
the measurement of damping timescale is reliable when it is larger than the typical cadence of LC.
We use the following criteria to select the reliable measurements of the damping timescale:
(1) The length of the LC should be ten times of the timescale at least;
(2) The derived timescale should be larger than the mean cadence of the LC. 
It is found that all the $\gamma$-ray characteristic timescales for the 23 sources are reliable.

The LCs are fitted in the observed frame, and the values of the damping timescales in Table \ref{tab:Posterior Parameters} are in the observed frame.
In order to get the timescale in the rest frame ($\tau^{\rm rest}_{\rm damping}$), 
the timescale should be corrected by cosmological
time dilation and and Doppler beaming effect, 
\begin{equation}\label{resttau}
\tau^{\rm rest}_{\rm damping}=\frac{\tau_{\rm DRW}\ \delta_{\rm D}}{1+z}\;,
\end{equation}
where $\delta_{\rm D}$ is the Doppler factor.
The Doppler factor for the $\gamma$-ray emission region in AGN jet is difficult to measure.
It is estimated by different methods based on, for example, modeling the broadband spectral energy distributions \citep{2018ApJS..235...39C,2020PASA...37...43P}, 
opacity of the $\gamma$-rays, and the brightness temperature of radio flare \citep{2017MNRAS.466.4625L}.
In Table \ref{tab:Doppler factor}, we list the average values of $\delta_{\rm D}$ for different types of AGNs estimated by recent three papers \citep{2017MNRAS.466.4625L,2018ApJS..235...39C,2020PASA...37...43P}.
One can see that the uncertainties on the Doppler factor are very large.
We use the average results of the three papers to correct the timescale.
The average $\delta_{\rm D}$ for blazars is 10.
It is found that $\tau^{\rm rest}_{\rm damping}$ is between 100 days and 1500 days, and the average $\tau^{\rm rest}_{\rm damping}$ is $\approx510\ $days.

\begin{figure}
    \centering
    {\includegraphics[width=12cm]{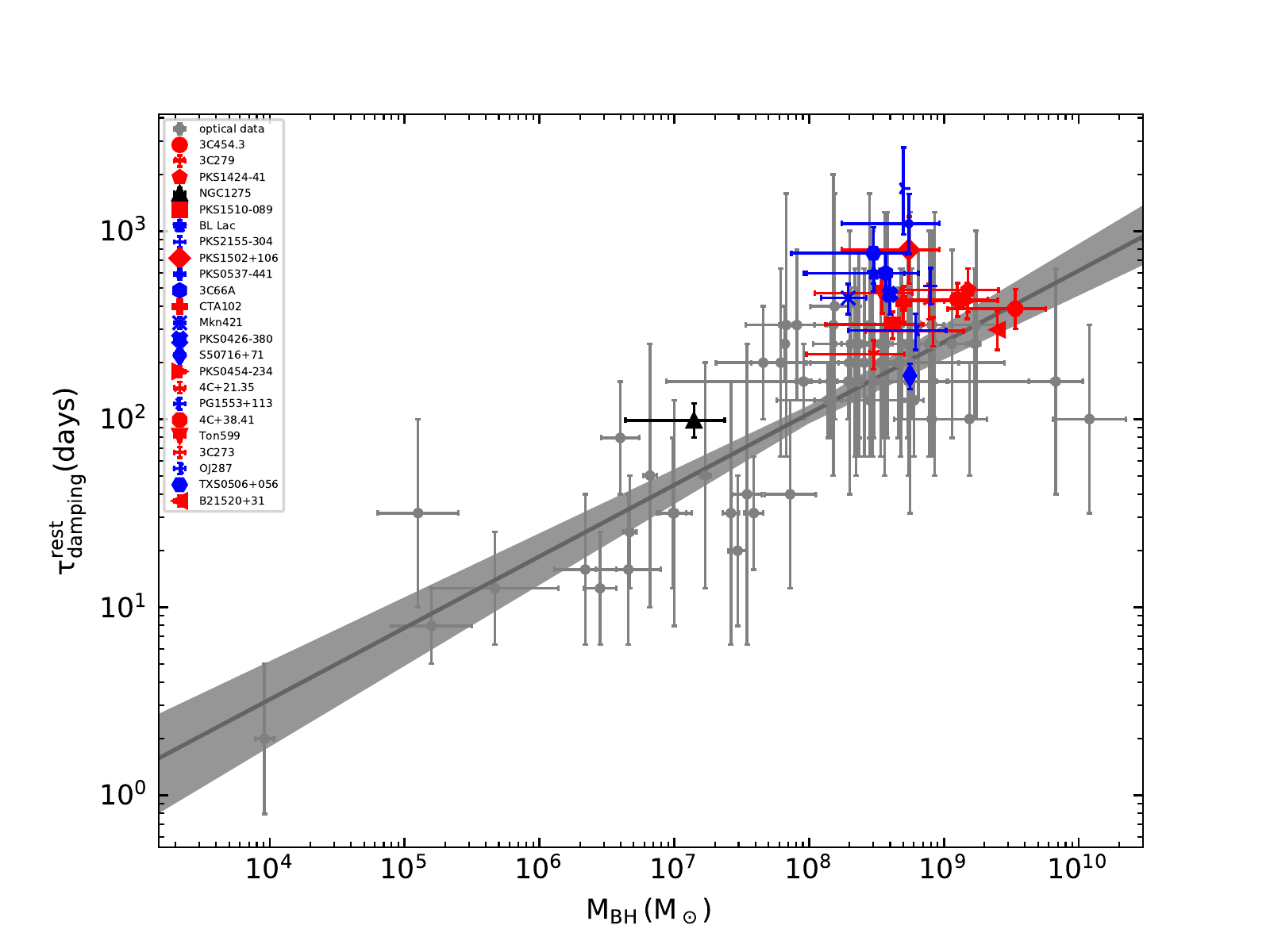}}
    \caption{Variability damping timescale (in the rest frame) as a function of black hole mass. 
    The gray data, lines and area represent optical results taken from \cite{2021Sci...373..789B}. 
    The data in color are our results from the $\gamma$-ray LCs of AGNs. Data points in red, blue and black represent FSRQ, BLL and RDG, respectively. 
    \label{fig:tau-mass}}
\end{figure}

The characteristic timescale in the optical variability of AGN accretion disk 
has been extensively studied \citep[e.g.,][]{2001ApJ...555..775C,2009ApJ...698..895K,2010ApJ...721.1014M,2016A&A...585A.129S,2021ApJ...907...96S}.
Recently, \citet{2021Sci...373..789B} reported a correlation between the optical characteristic timescale and the black hole mass.
We also show our results in the plot of $\tau^{\rm rest}_{\rm damping}-M_{\rm BH}$ (Figure~\ref{fig:tau-mass}), together with the results in \citet{2021Sci...373..789B}. 
It can be seen that the $\gamma$-ray variability timescales of AGNs occupy the same space with the optical variability timescales.
Namely, in the same range of the black hole mass, the $\gamma$-ray variability timescales are consistent with the optical variability timescales within the errors.
There is no correlation between the $\gamma$-ray variability timescale of AGN and the black hole mass.
This is probably due to the small dynamic range in the black hole mass in the sample of $\sim 10^{8}-10^{9} M_{\rm \odot}$ (with the exception of NGC 1275) making any correlation difficult to be identified. 
We have searched the $\gamma$-ray AGNs with smaller black hole mass. However, they are not bright enough to perform the variability analysis.
In Appendix~\ref{crab}, we use the $\gamma$-ray flares from Crab Nebula to extend the mass of the central engine to much smaller range.

The $\gamma$-ray $\tau^{\rm rest}_{\rm damping}$ values slightly and systematically lie above the optical relation of \cite{2021Sci...373..789B}.
In addition, we fit our $\gamma$-ray results and the optical results together, 
resulting in the best-fit relation, 
\begin{equation}\label{fittingmodel}
\tau^{\rm rest}_{\rm damping}=154.22^{+14.76}_{-15.75}(\frac{M_{\rm{BH}}}{10^{8}M_{\rm{\odot}}})^{0.43^{+0.04}_{-0.04}}\;,
\end{equation}
with an intrinsic scatter of \textbf{$0.23\pm 0.03$} dex and Pearson correlation coefficient \textbf{$r = 0.80$}. 
It is similar to the optical result.

\begin{figure}
    \centering
    {\includegraphics[width=12cm]{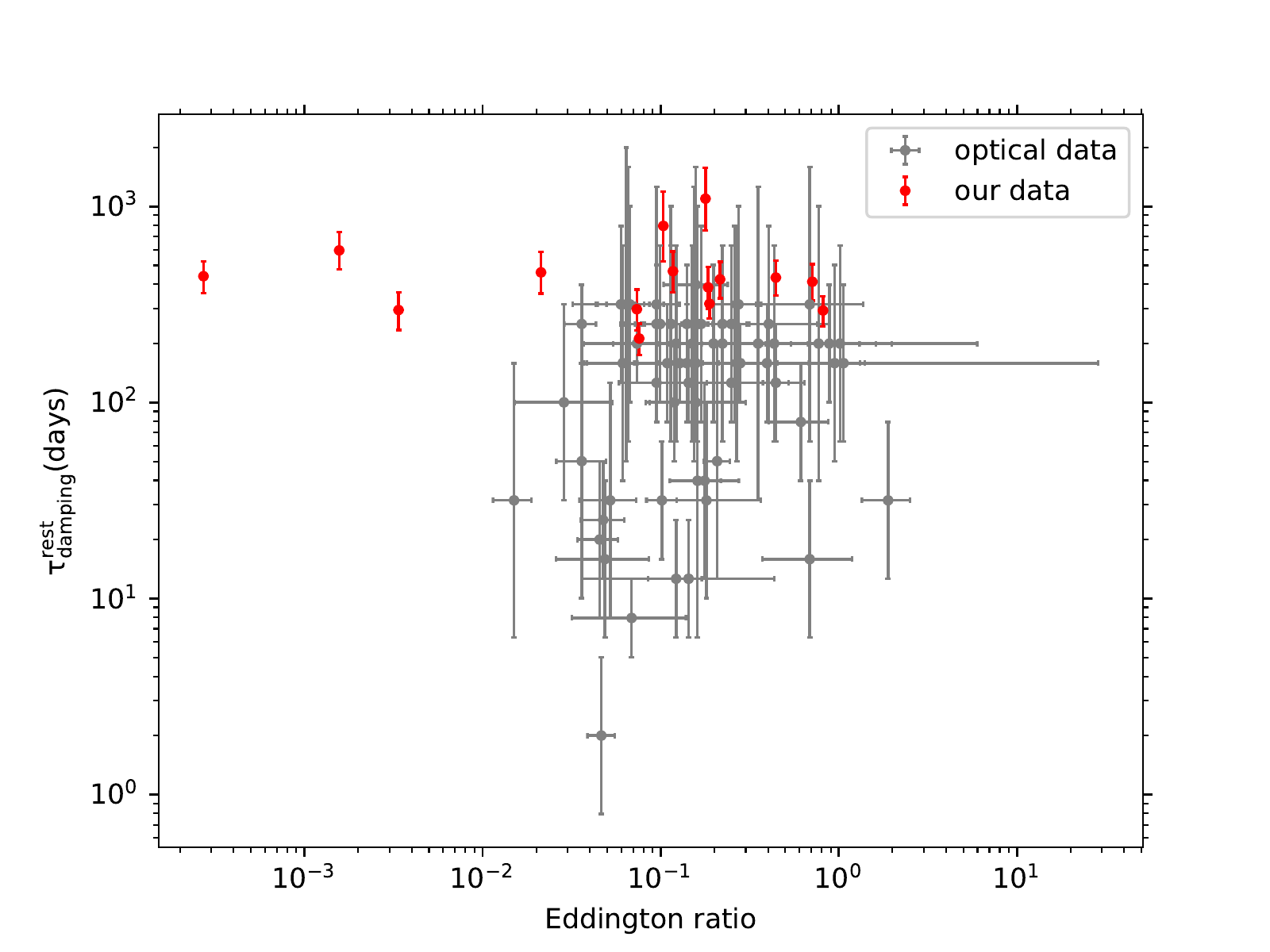}}
    \caption{Plot of the rest-frame timescale versus the Eddington ratio. 
    The gray data points with error bars are optical results taken from \cite{2021Sci...373..789B}. 
    The data points in red denote the $\gamma$-ray results of blazars in our samples. 
    \label{fig:eddington ratio}}
\end{figure}

We get the Eddington ratio (the ratio between accretion disk luminosity and the Eddington luminosity) for 15 out of 23 AGNs from \citet{2015MNRAS.450.3568X}, which are listed in Table~\ref{tab:information}. 
Except for Mrk 421, BL Lac and OJ 287, the rest of 12 sources have Eddington ratio between 0.01 and 1.
From Figure~\ref{fig:eddington ratio}, one can see that  the 12 sources have similar Eddington ratio with the normal quasars in \cite{2021Sci...373..789B}, 
and the characteristic timescale is independent on the Eddington ratio. 

\begin{deluxetable*}{ccccc}
	\tablecaption{Doppler factor for AGNs \label{tab:Doppler factor}}
	\tablewidth{20pt}
	\setlength{\tabcolsep}{5mm}{
	\tablehead{
	\colhead{FSRQ} & \colhead{BL Lac} & \colhead{RDG} & \colhead{Ref.} 
	}
	\startdata
	  11 & 14 & $\cdots$ & 1 \\
	15$\pm9$ & 9$\pm8$ & 1.4$\pm0.8$ & 2 \\
	 7$\pm4$ & 4$\pm3$ & $\cdots$ & 3
	\enddata 
	\tablecomments{Doppler factors with the errors for different types of AGNs taken from the references: 
	(1) \cite{2018ApJS..235...39C}, (2) \cite{2017MNRAS.466.4625L}, and (3) \cite{2020PASA...37...43P}. 
	}
	}
\end{deluxetable*}

\section{Discussion} \label{sec:Discussion} 

The stochastic process models are increasingly used to model $\gamma$-ray variability of blazars, 
providing an effective tool to study the statistical properties of the variability. 
We model the $\gamma$-ray variability of 23 bright LAT AGNs with the DRW and SHO models by performing the {\it celerite} package.
It is found that the DRW model with two parameters can fit the $\gamma$-ray variability of 23 AGNs successfully, 
and the fits with the SHO model with three parameters for the 23 sources are not improved.
The SHO is constrained in the over-damped mode ($Q<0.5$), and the strong degeneracy between $\omega_0$ and $Q$ leads to poor constraints on the two parameters.
The PSDs of the 23 sources are typical DRW form with $f_{\rm b}$ between $0.001\ \rm day^{-1}$ and $0.01\ \rm day^{-1}$. 

The brightest flares ($\gtrsim 10^{-5}\rm\ ph\ cm^{-2}\ s^{-1}$)  in 3C 279 and 3C 454.3 are poorly fitted by both DRW and SHO models. 
 Indeed, the extreme $\gamma$-ray flares in 3C 279 seem special \citep[e.g.,][]{2020NatCo..11.4176S}. 
 The $\gamma$-ray photon index in the extreme flare on Dec 16 2012 is 1.7 \citep{2015ApJ...807...79H}, 
 significantly smaller than the typical $\gamma$-ray photon index of 2.4 for FSRQ \citep{2020ApJS..247...33A}. 
 Minute-scale GeV $\gamma$-ray variability from 3C 279 was observed in an extreme flare on June 15 2016 \citep{2016ApJ...824L..20A}.
\citet{2013MNRAS.430.1324N} studied the individual $\gamma$-ray flares of 3C 454.3 with the flux above $0.71\times10^{-5}\rm\ ph\ cm^{-2}\ s^{-1}$, and found that the $\gamma$-ray flares of 3C 454.3 have more complex light curves than other blazars.
The extreme flares may have a different physical mechanism than the underlying long-term stochastic variability. 

These brightest flares are expected to have impact on the slope of PSD at high frequencies \citep{2019ApJ...885...12R}.
We have examined that the brightest flares in 3C 279 and 3C 454.3 cannot affect the modeling results for the long-term variabilities.
A further and careful study on the brightest flares is worthy of performing by  using an adaptive binning
algorithm.

The theoretical PSD expected by the one-zone leptonic emission model has been investigated \citep{2014ApJ...791...21F,2015ApJ...809...85F,2016MNRAS.458.3260C}.
\citet{2022ApJ...925..177T} simulated
multi-wavelength variability of blazars from a purely
numerical approach by using a time-dependent one-zone leptonic emission model.
They showed that a power-law PSD for the emission variability 
is produced by introducing stochastic variations for model parameters in the emission region, 
and the PSD is similar to the underlying power law of the model parameter variation.
No spectral break is found in their produced PSDs.
The results of \citet{2022ApJ...925..177T} indicate that in the frame of one-zone emission model, 
the physical processes associated with electron cooling, light crossing, and electron escape would not produce a break in the PSD. 
The broken frequencies we obtained are between $10^{-8}\ $Hz and $10^{-7}\ $Hz.
The corresponding intrinsic timescale is several hundred days at least, 
which cannot be the timescale corresponding to electron cooling or acceleration process.

The $\gamma$-ray timescales of AGNs we obtained are very close to the optical timescales obtained from modeling AGN accretion disk emissions in \cite{2021Sci...373..789B}.
\cite{2021Sci...373..789B} speculated that the optical timescales could be associated with the thermal timescales\footnote{The thermal timescale reads $t_{\rm th}=4.6\times\left(\frac\alpha{0.01}\right)^{-1}\left(\frac{M_{\rm BH}}{10^8M_{\rm \odot}}\right)\left(\frac R{100R_{\rm S}}\right)^{3/2}\ \ \rm yrs$, where $R$ is the emission distance on the accretion disk from the central black hole, $R_{\rm S}=2GM_{\rm BH}/c^2$ is the Schwarzschild radius, and $\alpha$ is the standard disk viscosity parameter.}
expected in the AGN standard accretion disk theory, and the optical variability may be driven by the thermal instability of the accretion disk.
The similarity between the $\gamma$-ray and optical characteristic timescales could imply a connection between jet and accretion disk.
The thermal instability may also causes the $\gamma$-ray variability in the jet.
However, the detailed mechanism that connects the accretion disk and the jet is unclear.
The $\gamma$-ray timescales are slightly larger than the optical timescales of normal quasars. 
This may be due to that the distance from the $\gamma$-ray emission region to accretion disk extends the intrinsic timescale from accretion disk.

\citet{2012ApJ...760...51R} modeled the nonthermal optical variabilities of  51 $\gamma$-ray blazars, 
and found that blazar optical $\tau^{\rm rest}_{\rm damping}$ peaks at $\sim$1000 days (assuming a typical Doppler factor of 10), 
which is systematically larger than the $\gamma$-ray $\tau^{\rm rest}_{\rm damping}$ in this work.
The discrepancy between blazar $\gamma$-ray and optical $\tau^{\rm rest}_{\rm damping}$ 
may imply that the $\gamma$-ray and optical emissions are produced in different regions. 
The $\gamma$-ray emission region is closer to the accretion disk than the optical emission region.
\citet{2012ApJ...760...51R} found that blazar nonthermal optical characteristic timescales are $\sim$4 times smaller than normal quasars. 
They considered that the discrepancy between the optical characteristic timescales for blazars and normal quasars could be caused by the Doppler effect, 
if the jet variability and accretion disk variability have the same origin. 
Combining with our $\gamma$-ray results, we suppose that the discrepancy between the characteristic timescales for blazars and normal quasars is not only caused by the Doppler effect, 
but also related to the location of the jet emission region (the distance from the accretion disk).
 The jet long-term variability may be the convolution of the accretion disk variability with a transfer function which is related to 
 Doppler factor and the distance from the jet emission region to the accretion disk at least.

\section{Summary} \label{sec:Summary}

We have applied a stochastic process method to the $\sim$12.7 yr Fermi-LAT LCs of 23 jetted AGNs in order to investigate the $\gamma$-ray variability properties.
The SHO and DRW models are both used to model the long-term LCs.
Our main results are as follows.

$(\romannumeral1)$ The long-term variability of 23 sources in our sample can be described well by both SHO and DRW models. 
However, the modelings with the SHO are not improved, and the parameters $\omega_0$ and $Q$ are poorly constrained. 
This suggests that the DRW model is preferred over the SHO model for the $\gamma$-ray long-term variability of AGNs.
The PSDs for the 23 sources are the typical DRW PSD form.

$(\romannumeral2)$ The intrinsic characteristic timescale of AGNs extracted from modeling the $\gamma$-ray variability is between 100 days to 1500 days.
Such a long timescale cannot be produced in a one-zone leptonic emission model within the typical parameter space.
In the plot of $\tau^{\rm rest}_{\rm damping}-M_{\rm BH}$, the $\gamma$-ray timescales obtained from jet emissions occupy almost the same space with the optical timescales obtained from the accretion disk emissions.
Both the $\gamma$-ray and optical timescales are consistent with the thermal timescale expected by the AGN standard accretion disk.
It may indicate a connection between the jet and the accretion disk.

In conclusion, our results suggest that the origin of the $\gamma$-ray variability could be related to the thermal instability in the accretion disk, however the detailed process that drives the variability is unclear.

\acknowledgments
We thank the referee for valuable suggestions and Dr. Xiaoyuan Huang (PMO) for providing the $\gamma$-ray flare data of Crab Nebula.
This work is partially supported by National Key R \& D Program of China under grant No. 2018YFA0404204, 
and the National Natural Science Foundation of China (U1738211 and 11803081). 
H. Y. Zhang acknowledges the financial support from Scientific Research Fund project of Yunnan Education Department (2022Y053) and Graduate Research innovation project of Yunnan University (2021Y034).
The work of D. H. Yan is also supported by the CAS Youth
Innovation Promotion Association and Basic research Program of Yunnan Province (202001AW070013).

{\it Facility:} Fermi(LAT)

{\it Software:} Fermitools-conda, celerite \citep{2017AJ....154..220F}, emcee \citep{2013PASP..125..306F}, NumPy \citep{2020NumPy-Array}, Matplotlib \citep{2007CSE.....9...90H}, Astropy \citep{2013A&A...558A..33A,2018AJ....156..123A}, SciPy \citep{2020SciPy-NMeth}.

\appendix

\section{Crab Nebula $\gamma$-ray flare}
\label{crab}

\begin{figure}
    \centering
    {\includegraphics[width=12cm]{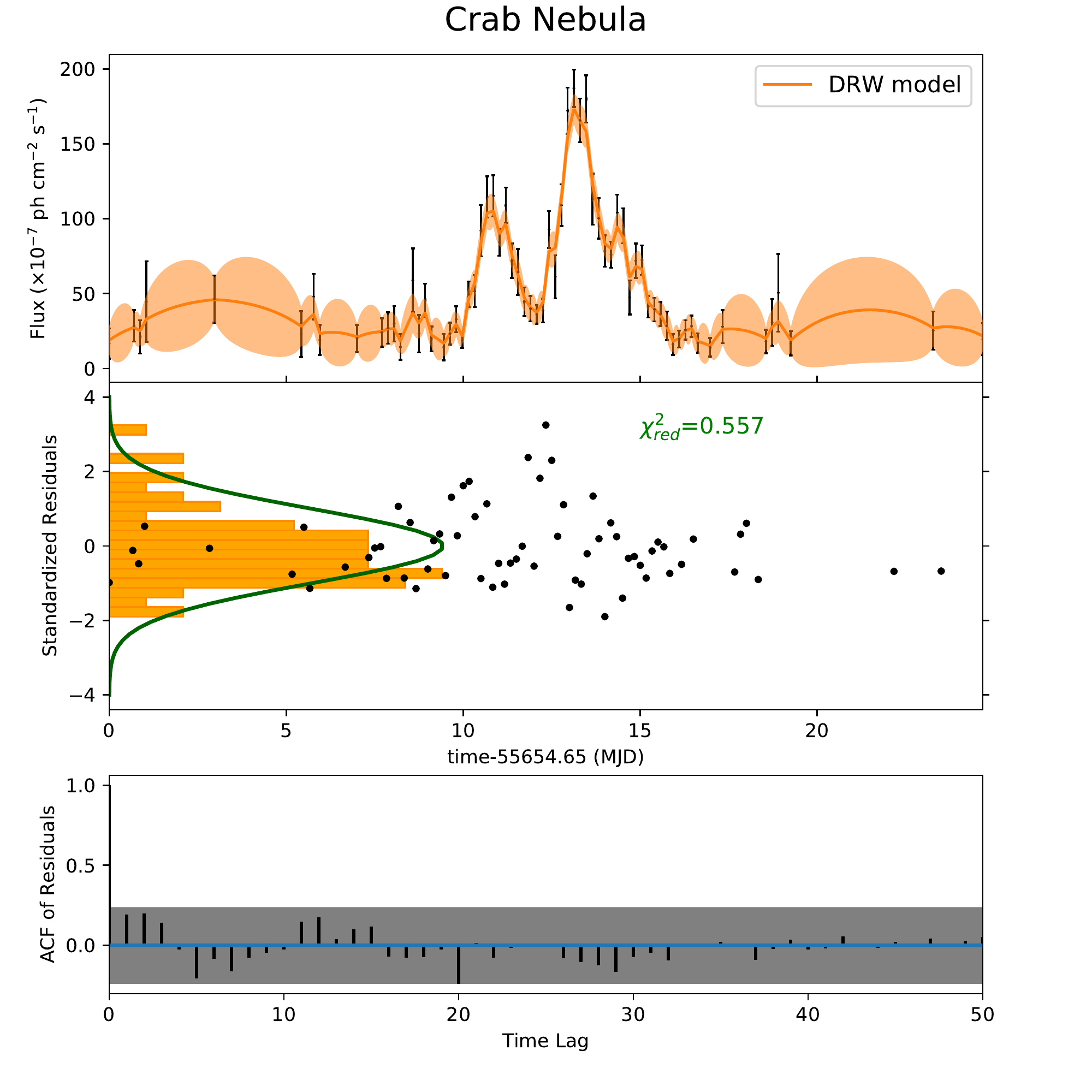}}
    \caption{DRW modeling results of $\gamma$-ray flare (MJD 55654.65-55678.65) from the Crab Nebula. The symbols and lines are the same as those in Figure~\ref{fig:celerite fit}.
    \label{fig:crab_fit}}
\end{figure}

\begin{figure}
    \centering
    {\includegraphics[width=7cm]{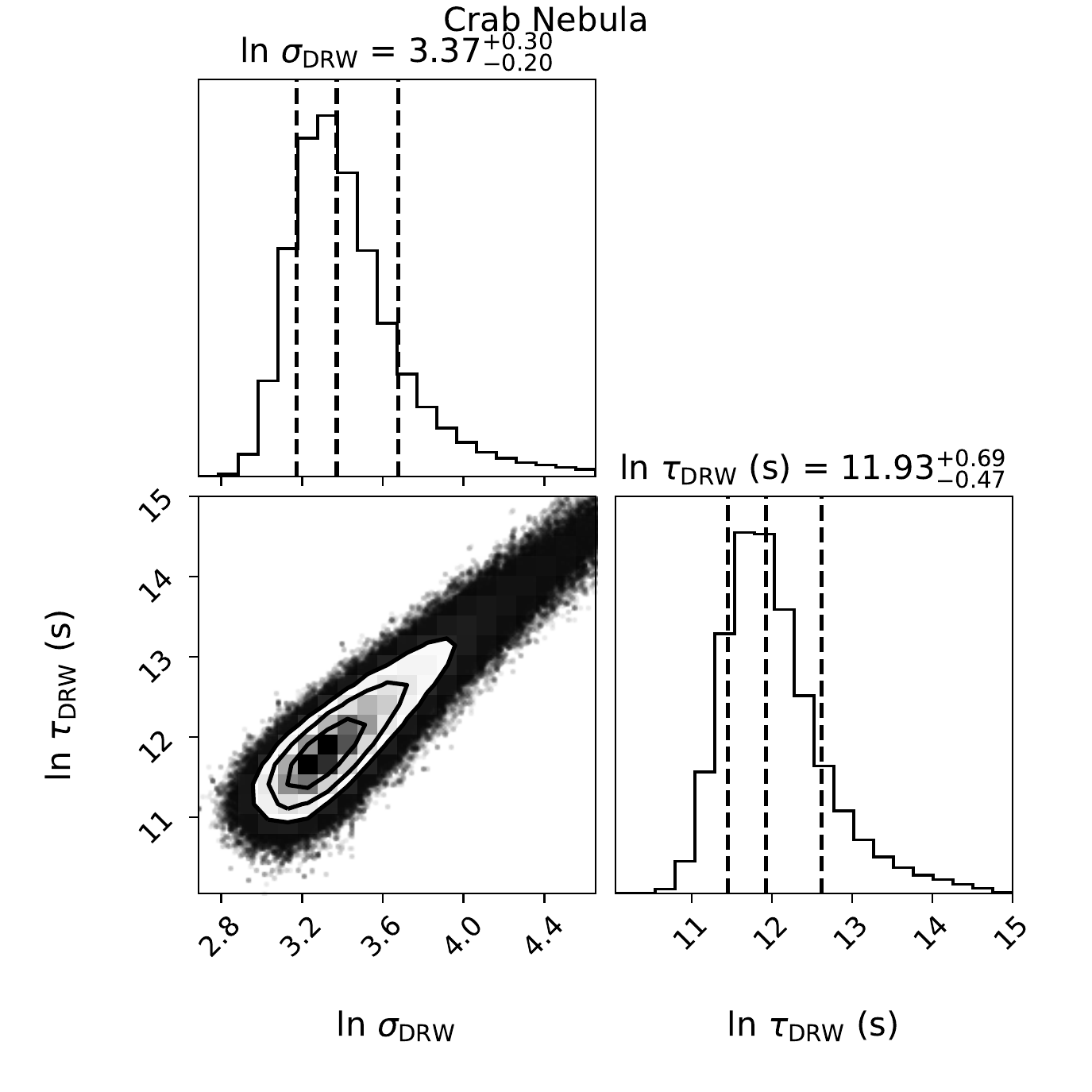}}
    {\includegraphics[width=9cm]{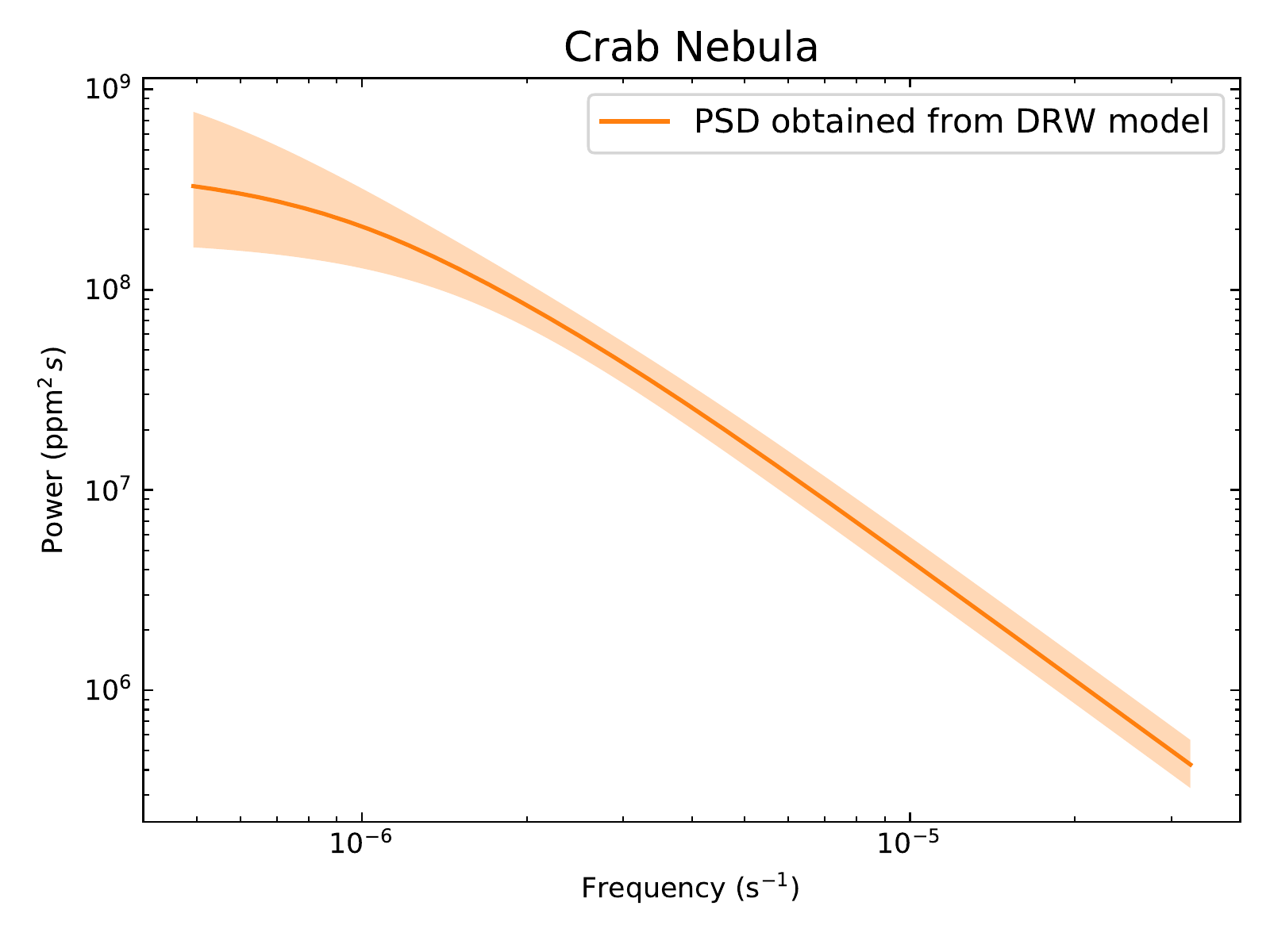}}
    \caption{Left panel: posterior probability densities of DRW parameters for the Crab Nebula. 
    The symbols and lines are the same as those in Figure~\ref{fig:SHOparam distribution}. 
   Right panel: DRW PSD of the $\gamma$-ray LC of the Crab Nebula. 
    The corresponding color region denotes 1$\sigma$ confidence interval.
    \label{fig:crab_psdparam}}
\end{figure}

\begin{figure}
    \centering
    {\includegraphics[width=12cm]{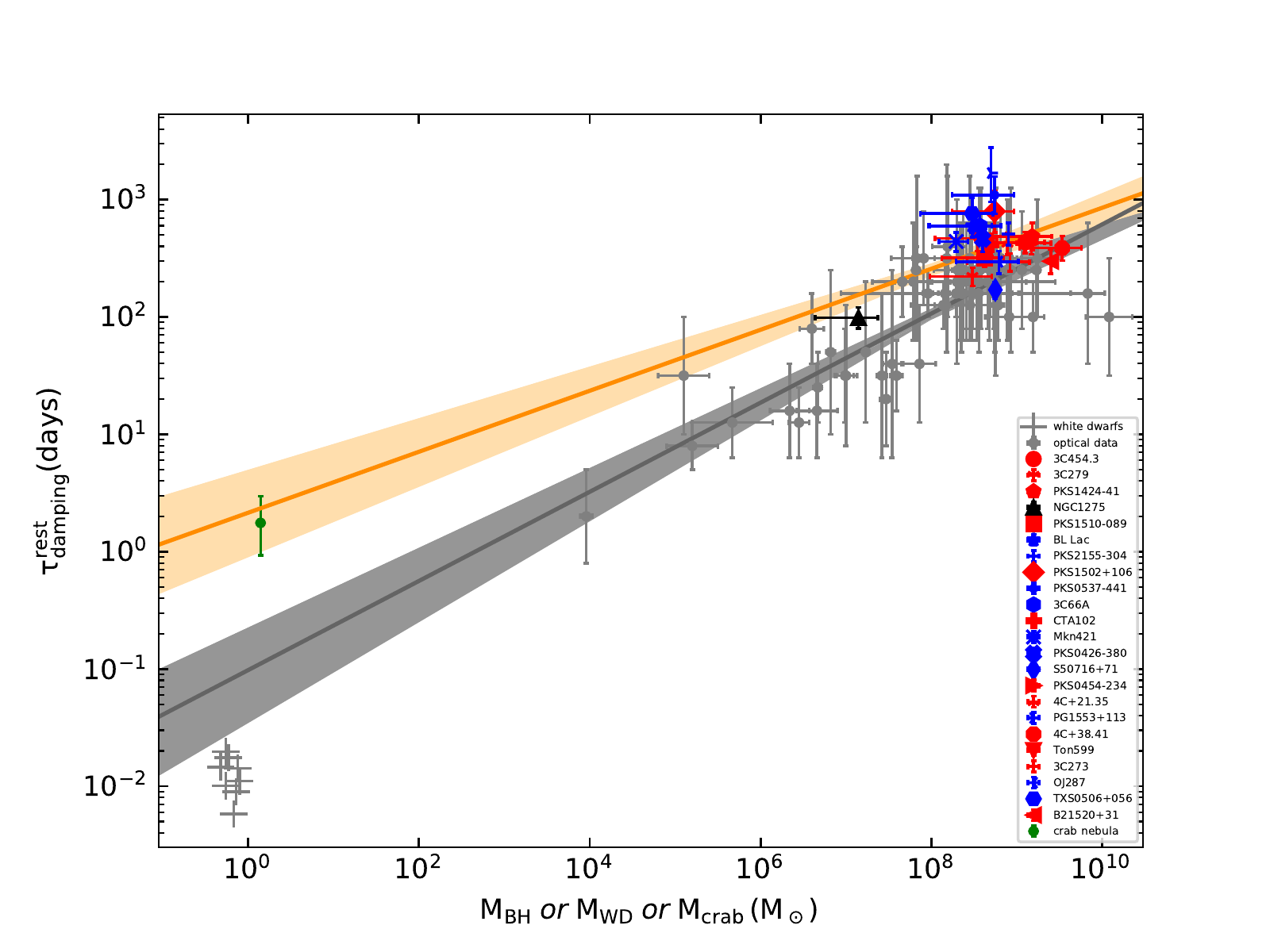}}
    \caption{Variability damping timescale (in the rest frame) as a function of the mass of the central engine. 
    The gray data, lines, area as well as the crosses represent optical results taken from \cite{2021Sci...373..789B}. 
    The data in color are our results from the $\gamma$-ray LCs of AGNs and the Crab Nebula. 
    The orange line and shaded band are the best-fit relation and 1$\sigma$ uncertainty for 23 AGNs and the Crab Nebula.
    \label{fig:tau-mass2}}
\end{figure}

GeV $\gamma$-ray flares from the Crab Nebula were observed by AGILE \citep{2011Sci...331..736T} and Fermi-LAT \citep{2011Sci...331..739A}.
The  central pulsar, PSR B0531+21, has a mass of 1.4 solar mass.
To extend the $\gamma$-ray $\tau^{\rm rest}_{\rm damping}$-mass relation to much smaller mass, we consider the $\gamma$-ray flares from the Crab Nebula.
\cite{2021ApJ...908...65H} identified 17 flares in the $\gamma$-ray emission from Crab Nebula.
The flare during MJD 55654.65-55678.65 has good sample and the flux uncertainties are relatively small.
We use the DRW model to model 4-hr binning LC during MJD 55654.65-55678.65. The fitting results, PSD and posterior probability densities of parameters are shown in Figure~\ref{fig:crab_fit} and Figure~\ref{fig:crab_psdparam}.
We obtain the characteristic timescale $1.8^{+1.2}_{-0.8}$ days. This timescale is less than 1/10 of the length of the LC and larger than the mean cadence (0.36 days), which is reliable.
We use the result to extend the $\gamma$-ray $\tau^{\rm rest}_{\rm damping}$-mass relation (Figure~\ref{fig:tau-mass2}).
There is a correlation (Pearson correlation coefficient $r=0.90$) between the $\gamma$-ray characteristic timescale and mass when adding the result of Crab Nebula, i.e.,
\begin{equation}\label{gamma-fitmodel}
\tau^{\rm rest}_{\rm damping}=257.52^{+28.49}_{-33.21}(\frac{M_{\rm{BH}}}{10^{8}M_{\rm{\odot}}})^{0.26^{+0.04}_{-0.04}}\;,
\end{equation}
with an intrinsic scatter of \textbf{$0.21\pm 0.05$} dex.

\bibliography{ms.bib}{}
\bibliographystyle{aasjournal}

\end{CJK*}
\end{document}